\documentclass[aps,prl,twocolumn,nofootinbib,superscriptaddress]{revtex4-2}
\pdfoutput=1
\usepackage{graphicx}
\usepackage{dcolumn}   
\usepackage{bm}       
\usepackage{amsmath,amssymb,amsxtra,bm,epsfig}   
\usepackage{float}
\usepackage[dvipsnames]{xcolor}
\usepackage{xspace}
\usepackage{hyperref}
\usepackage{cancel,soul,ulem}
\usepackage{orcidlink}  
\hypersetup{
	colorlinks=true,
	linkcolor=red,
	filecolor=magenta,
	urlcolor=JungleGreen,
	citecolor=RoyalBlue}

\newcommand{\bmt}{\begin{pmatrix}}
	\newcommand{\emt}{\end{pmatrix}}
\newcommand{\ba}{\begin{array}{c}}
	\newcommand{\ea}{\end{array}}
\newcommand{\be}{\begin{equation}}
	\newcommand{\ee}{\end{equation}}
\newcommand{\bea}{\begin{eqnarray}}
	\newcommand{\eea}{\end{eqnarray}}

\newcommand{\bi}{\begin{itemize}}
	\newcommand{\ei}{\end{itemize}}
\newcommand{\baz}{\begin{array}{cc}}
\newcommand{\besub}{\begin{subequations}}
\newcommand{\eesub}{\end{subequations}}

\newcommand{\stkout}[1]{\ifmmode\text{\sout{\ensuremath{#1}}}\else\sout{#1}\fi}

\begin{document}

	\title{Unraveling Freeze-in Dark matter through the echoes of gravitational waves}

	\author{Partha Konar\,\orcidlink{0000-0001-8796-1688}}
	\email{konar@prl.res.in}
	\affiliation{Theoretical Physics Division, Physical Research Laboratory, Ahmedabad, 380009, India}

	\author{Sudipta Show\,\orcidlink{0000-0003-0436-6483}}
	\email{sudiptas@iitk.ac.in}
	\affiliation{Department of Physics, Indian Institute of Technology Kanpur, Kanpur 208016, India}

\begin{abstract}
In the quest to unravel the dark sector, feebly interacting freeze-in dark matter presents an intriguing possibility, plausibly explaining the consistent null results from various dark matter experiments. We propose a unique imprint in the form of gravitational waves generated during the freeze-in production of dark matter from heavy particle decay in the early universe. This characteristic gravitational wave signature can serve as a powerful probe for freeze-in dark matter. Our study indicates that future high-frequency gravitational wave experiments can detect these waves, offering a novel avenue to critically test the underlying conditions and requirements of this dark matter paradigm, which typically lie beyond the reach of current and planned dark matter detection experiments.			
\end{abstract}


\maketitle

\textit{Introduction}---Despite compelling evidence of its omnipresence primarily relying on gravitational interactions, the dark matter (DM) puzzle remains one of the most significant challenges in modern physics. The Weakly Interacting Massive Particle (WIMP) paradigm, historically the most extensively studied and popular DM candidate \cite{Arcadi:2017kky, Roszkowski:2017nbc}, is currently under tension from numerous DM detection experiments operating at various energy and cosmic frontiers. Consequently, numerous exciting mechanisms have been proposed, garnering significant interest in the ongoing effort to understand the dark sector. Notably, DM as a Feebly Interacting Massive Particle (FIMP) \cite{Hall:2009bx, Bernal:2017kxu} offers a natural explanation for the observed null results thus far due to the inherently feeble nature of its interactions with Standard Model (SM) particles. While this offers an appealing choice, it comes at the expense of detectability, making direct \cite{LZ:2022lsv, PandaX:2024qfu, XENON:2025vwd} and indirect \cite{Fermi-LAT:2015att, Fermi-LAT:2016uux} detection experiments nearly impossible probes. Alternatively, direct production of other heavier states in an enriched dark sector can provide a limited but exciting scope in collider experiments. In this letter, we present a complementary strategy to probe such DM through gravitational waves (GWs) generated during their freeze-in production via the decay of heavy states.
	 
In a minimalistic setup, Higgs ($H$) portal scalar DM ($\chi$) with a tiny interaction strength, characterized by the coupling $y H^2 \chi^2$, generally struggles to produce observable signatures at colliders. In contrast, a broader class of models featuring the interaction term $y X \xi \chi$ can yield unconventional collider signatures. Here, $X$ represents a Beyond Standard Model (BSM) mediator, $\xi$ is an SM particle, and $\chi$ denotes the DM candidate. Such a DM can be generated in the early universe through the gradual decay of the heavy particle $X$ in thermal equilibrium, with its final abundance determined by the freeze-in mechanism~\cite{Hall:2009bx, Bernal:2017kxu}. The smallness of the freeze-in coupling $y$ ensures that the DM remains non-thermalized with the thermal bath throughout the evolution of the universe. Due to its considerable interaction with SM, $X$ can now be produced abundantly at the collider experiment. Its subsequent decay into DM can potentially lead to unique signatures depending on the specific model. Such DM can be probed at the Large Hadron Collider (LHC) by exploring unique signatures with displaced vertex (DVs)~\cite{ATLAS:2017tny}, long-lived particle searches (LLPs)~\cite{Curtin:2018mvb}.

The specific nature of DM defines the characteristics of the BSM and SM particles involved in its interactions. Many intriguing models have been previously explored based on the $y X\xi\chi$ interaction topology. These models can be categorized as as hadrophilic DM models~\cite{Belanger:2018sti, Garny:2018icg, DEramo:2017ecx, Ghosh:2024nkj, Becker:2023tvd},  leptophilic DM models~\cite{Belanger:2018sti, Chakraborti:2019ohe, DEramo:2017ecx, Becker:2023tvd}, singlet-doublet Dirac~\cite{No:2019gvl, Ghosh:2021wrk, Das:2023owa} and Majorana DM models~\cite{Calibbi:2015nha, Calibbi:2018fqf}. The characteristics of these models are outlined in the Supplemental Material.

Each of the models detailed above has been investigated within the context of the freeze-in mechanism, demonstrating that such models can be probed at colliders through DVs or LLPs, at least for a suitable freeze-in coupling strength.
Furthermore, specific constituents dominating the cosmic background of the early universe can significantly influence DM production, potentially demanding a revised coupling strength. For example, a popular choice of modified cosmology, such as a fast-expanding universe, requires a larger interaction strength than a radiation-dominated universe to be consistent with the relic density constraint. 
This effectively alters the collider search strategy. For instance, as shown in references~\cite{Das:2023owa, Ghosh:2024nkj}, DVs no longer serve as an adequate search tool; instead, boosted jet techniques can be instrumental for probing such scenarios due to the accelerated decay necessitated by non-standard cosmology. It is important to note that collider searches primarily provide a lower bound on the heavy particle mass, which is dependent on the specific microscopic model.

Probing freeze-in DM via DVs at colliders necessitates that heavy BSM states decay within the tracker. Simultaneous fulfillment of this requirement and the observed DM relic abundance typically constrains the DM mass at the order of keV~\cite {Calibbi:2018fqf}\footnote{Notably, Lyman-$\alpha$ flux-power spectra data imposes a lower bound on the DM mass, which is around 5.3 keV, by constraining the free-streaming of DM~\cite{Irsic:2017ixq}. However, the most stringent lower mass bound for freeze-in dark matter is around 15.67 keV~\cite{DEramo:2020gpr}.}. Interestingly, the warm DM~\cite{Viel:2013fqw} within this characteristic mass range can not only be explored through astrophysical experiments but also directly addresses the long-standing small-scale problems of $\Lambda$CDM (standard picture of cosmology)~\cite{Lovell_2012, Bode:2000gq} by suppressing the structure formation at small scale.

While collider searches can only offer a very narrow vantage point, they cannot fully illuminate the profound mystery of freeze-in DM scenarios. In practical terms, the prospect remains bleak for any present or upcoming DM detection experiments. In this letter, we advocate an independent and unique signature of it in the form of the emission of GWs that can occur from the decay of heavy particles in the early universe, an unavoidable consequence of the minimal coupling between the SM and BSM particles to gravity. This signature, once produced, propagates undisrupted through cosmic history and can be detected as a new tool to investigate freeze-in DM processes occurred in the ``baby universe''. GW generation can occur during DM production through both scattering and decay processes, depending on which contributes dominantly in a given model. For a simple demonstration, we discuss GW production via decay within a model, particularly model I or II, where DM ($\chi$) is a scalar, the heavy particle is a vector-like fermion ($X$), and the SM particles involved in this interaction are SM fermions $\xi$.

\textit{Graviton bremsstrahlung}---Gravitons couple to the energy-momentum tensor. Consequently, at the lowest order, a single graviton can be emitted from any of the interacting fields, as well as from the interaction vertex itself. In this framework, the action describing the interaction of the graviton with SM and BSM fields, derived from the Einstein-Hilbert action, is expressed as:
$S\supset\int d^4x\sqrt{-g} [\frac{M_P^2}{2}\mathcal{R}+\mathcal{L}_{\text{SM}}+\mathcal{L}_{\text{DM}}+\mathcal{L}_X],$
where $g$ denotes the determinant of the metric $g_{\mu\nu}$, $M_P(=2.4\times 10^{18}~\text{GeV})$ corresponds	 to the reduced Planck mass and $\mathcal{R}$ refers to the Ricci scalar. Furthermore,  $\mathcal{L}_{\text{SM}}$ signifies Lagrangian of the SM, while  $\mathcal{L}_{\text{DM}}$, $\mathcal{L}_X$ denote the Lagrangian for the DM and $X$ respectively, encompassing both their mass and kinetic terms. Applying the weak field approximation, $g_{\mu\nu}$ can be expanded around the Minkowski metric $\eta_{\mu\nu}(=diag(1,-1,-1,-1))$ as, $g_{\mu\nu}(\simeq\eta_{\mu\nu}+(2/M_P) h_{\mu\nu}+..)$.  Now, retaining the interaction terms only to the first order in $2/M_P$, one  obtains the following interaction term between the canonically normalized graviton $(h_{\mu\nu})$ and the stress-energy tensor $T_i^{\mu\nu}$ for any particle $i$, which denotes SM as well as BSM particles, as~\cite{Choi:1994ax, Holstein:2006bh}
$\mathcal{L}_{\text{int}}^{\text{grav}}\supset-\frac{2}{M_P}	h_{\mu\nu}\sum_iT_i^{\mu\nu}$.
In particular, the stress-energy tensors for fermion ($i=f$) and scalar ($i=\phi$) can be expressed as
\begin{align}\nonumber
	&T_f^{\mu\nu}=\frac{i}{4}[\bar f\gamma^\mu\partial^\nu f+\bar f\gamma^\nu\partial^\mu f]-\eta^{\mu\nu}\bigg[\frac{i}{2}\bar f\gamma^\sigma\partial_\sigma f -m_f \bar f f\bigg],\\
	& T_\phi^{\mu\nu}=\partial^\mu\phi\partial^\nu\phi-\eta^{\mu\nu}\bigg[\frac{1}{2}\partial^\sigma\phi\partial_\sigma\phi-V(\phi)\bigg],
\end{align}
where $V(\phi)$ corresponds to the potential of the scalar field $\phi$. The graviton production occurs via three-body decay of the heavy particle $X$, specifically via the graviton \textit{bremsstrahlung} without affecting the DM production via the 2-body decay of the same. Figure~\ref{GW_prod} depicts the corresponding Feynman diagrams.
\begin{figure}[t]
	\centering
	\includegraphics[width=4cm, height=1.5cm]{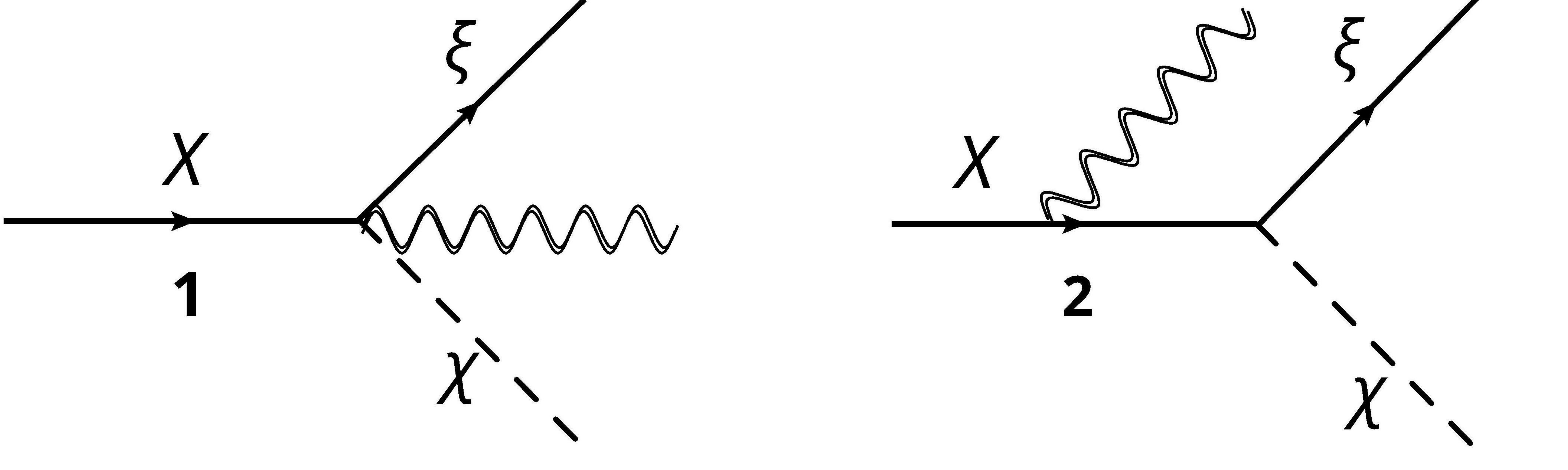}~
		\includegraphics[width=4cm, height=1.5cm]{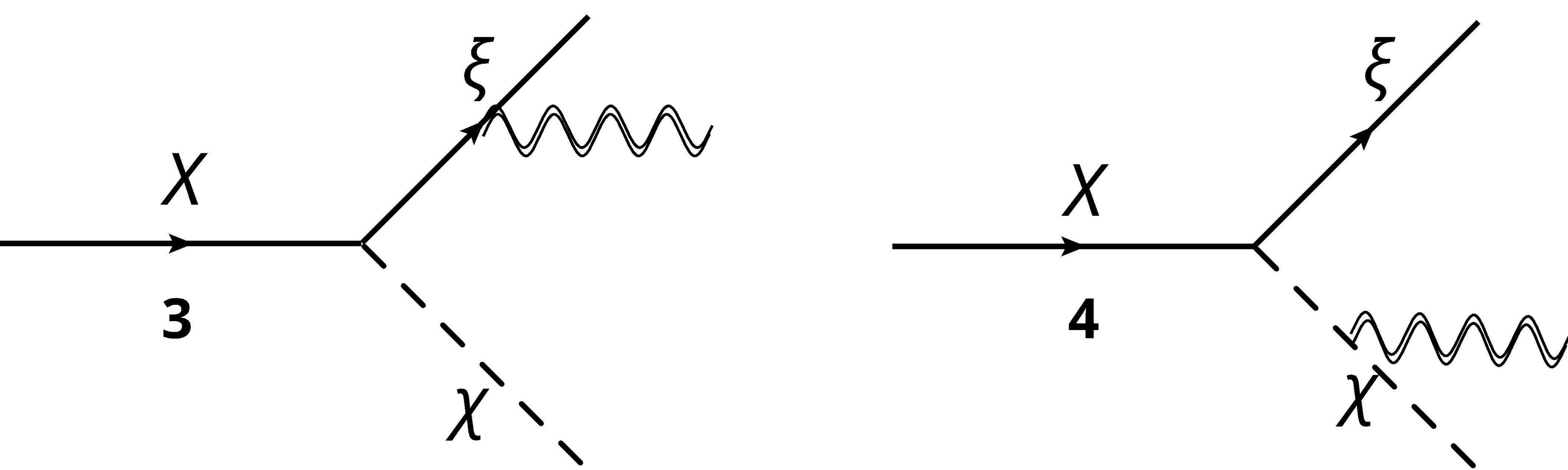} 
	\caption{ Feynman diagrams for graviton production.}
	\label{GW_prod}
\end{figure}
The amplitude associated with the first diagram is proportional to $\epsilon_{\mu\nu}\eta^{\mu\nu}$ ($\epsilon_{\mu\nu}$ denotes the polarization of the graviton), hence vanishes due to the traceless criterion of the massless graviton. Similarly, the contribution from the second diagram also vanishes since the amplitude is proportional to the momentum of the decaying particle ($\mathcal{M}_2\propto p_X^\mu p_X^\nu\epsilon_{\mu\nu}$) and the calculation is performed in the rest frame of the mother particle.
Interestingly, both the square of the amplitude of the third diagram and the interference between the third and the fourth diagram are proportional to the square of the mass of the daughter fermion ($|\mathcal{M}_3|^2, \mathcal{M}_3.\mathcal{M}_4^* \propto m_\xi^2$). We calculate the production of GWs in the very early universe, prior to electroweak symmetry breaking, when all SM particles are essentially massless. Consequently, in the massless limit of the daughter particles, both these contributions also approach zero.
Effectively, only the rightmost diagram contributes to graviton production, provided the masses of the product particles are negligibly small compared to the parent particle ($m_\chi, m_\xi\ll m_X$). Under this criterion, the differential decay width for the $1\rightarrow 3$ process with graviton \textit{bremsstrahlung} is given by
\begin{align}
	\frac{d\Gamma^{1\rightarrow 3}}{dE_{\text{gw}}}=\frac{y^2}{768\pi^3}\frac{m_X^2}{M_P^2}\mathcal{F}(x).
	\label{diff_decay_width}
\end{align}	
Here, $\mathcal{F}(x)=(2-x)(1-2x)^2/x$ with $x=E_{\text{gw}}/m_X$ and $E_{\text{gw}}(=2\pi f)$, the energy (frequency) of the graviton, spans over an range $0\le E_{\text{gw}}\le m_X/2$ (i.e. $0\le x\le 1/2$).

\textit{Dark matter production}---Before computing the GWs production during the decay of heavy particles, it is important to discuss the freeze-in production of DM from the same. This will enable a clearer illustration of how the underlying dynamics of freeze-in directly correlate with the resulting energy density of the emitted gravitons.
The number density ($n_{\text{DM}}$) of the DM produced via the decay of particle $X$ in thermal equilibrium can be traced by solving the following Boltzmann equation:
\begin{equation}
	\frac{dn_\chi}{dt}+3H n_\chi=\langle\Gamma^{1\rightarrow 2}\rangle n_X^{\text{eq}},
	\label{Boltz_nd}
\end{equation}
where, $H(T)(=\sqrt{\frac{{\pi^2 g_*}}{90}}\frac{T^2}{M_P})$ represents the Hubble rate during the radiation-dominated era,  $\langle\Gamma^{1\rightarrow 2}\rangle (=\Gamma^{1\rightarrow 2}\frac{K_1(m_X/T)}{K_2(m_X/T)})$ is the thermally averaged decay width, and $n_X^{\text{eq}}(=\frac{g_X m_X^2 T}{2\pi^2}K_2(m_X/T))$ is the equilibrium number density of the decaying particle. Here, $M_P(=2.4\times 10^{18}~\text{GeV})$, $g_*$ and $g_X$ represent the reduced Planck mass, relativistic degrees of freedom of the universe, and degrees of freedom of the particle $X$, respectively.
The two-body decay width, $\Gamma^{1\rightarrow 2}=({y^2}/{32\pi})m_X$ assumes a negligible mass for both the SM fermion  and the DM particle compared to $m_X$. Note that the annihilation of SM fermions can also produce dark matter from the 2-2 scattering via a heavy mediator, but the scattering cross-section is suppressed by $y^2$ compared to decay. So, one can safely neglect the contribution of scattering, which in turn dictates that GW emitted in the form of graviton \textit{bremsstrahlung} is also negligibly small compared to the same from decay. It is customary to rewrite Eq.(\ref{Boltz_nd}) in terms of abundance $Y(=n/s)$ and dimensionless variable $z(=m_X/T)$ where $s(=({2\pi^2}/{45})g_{*s}T^3)$ corresponds to entropy density with $g_{*s}$ being the entropic degrees of freedom. This reformulation yields:
\begin{equation}
	\frac{dY_\chi}{dz}=\frac{\langle\Gamma^{1\rightarrow 2}\rangle}{Hz}Y_X^{\text{eq}},
	\label{Boltz_abun}
\end{equation}
Once the final freeze-in abundance of DM is obtained numerically by solving Eq.~(\ref{Boltz_abun}), the DM relic density can be evaluated using the expression:\footnote{In principle, the Boltzmann equation should contain an extra term incorporating the change in $g_*$ which we have ignored since $g_*$ remains constant throughout the evolution of our interest.}
\begin{equation}
	\Omega h^2=2.755\times 10^8 m_\chi Y_\chi(z\rightarrow\infty),
	\label{relic_density}
\end{equation}
DM must satisfy the to satisfy the relic density constraint,$\Omega h^2=0.12$~\cite{Planck:2018vyg}. An analytical expression for the relic density can also be derived by solving Eq.~(\ref{Boltz_abun}):
\begin{equation}
	\Omega h^2={2.755\times 10^8}\frac{M_Pg_X\Gamma^{1\rightarrow 2}}{\sqrt{g_*}g_{*s}}\frac{m_\chi}{m_X^2}\frac{3\pi}{2}.
	\label{relic_ana}
\end{equation} 	

At this point, it is helpful to examine the evolution of the equilibrium abundance of $X$ ($Y_X^{\text{eq}}$) and the abundance of dark matter ($Y_\chi$) as functions of $x$, given a specific set of parameters $(m_\chi, m_X, y)$. For a detailed illustration, please refer to the Supplemental Material.

From Eq.~(\ref{relic_ana}), it is evident that for a fixed DM mass, the relic density constraint imposes the condition $y^2/m_X=\text{constant}$, We have verified that the DM never thermalizes for our choice of masses and required couplings. See Supplemental Material for further clarity.
This criterion leads to the saturation point ($z_s$) for freeze-in being fixed at approximately 30, irrespective of the value of $m_X$, due to the ($y^2\propto m_X$) relationship. If DM mass is larger than 12 keV, as we would examine further, the relic density constraint can be satisfied with a smaller final DM abundance ~(see Eq.~(\ref{relic_density})), thus requiring a smaller Yukawa coupling.

It is worth noting that DM production can also occur from the out-of-equilibrium decay of a heavy particle, a scenario distinct from the freeze-in process described above. In this case, the heavy particle decouples from the thermal bath while still relativistic, and its departure from thermal equilibrium occurs similar to the leptogenesis scenario~\cite{2006hep.ph....8347S, 2007hep.ph....3087C}. This requires simultaneously solving the evolution equations for both the heavy particle's number density and the DM's number density (Eq.~(\ref{Boltz_abun})). The heavy particle entirely decays into DM, resulting in a final DM abundance independent of the coupling strength. The coupling, however, determines the timescale for the heavy particle's complete decay into DM. In this specific context, we find that this scenario typically leads to an overabundant DM, rendering it unviable in the present setup.

\textit{Gravitational wave generation}---We now turn to the context of GW production via graviton \textit{bremsstrahlung} from the same decay process. As illustrated in Figure~\ref{GW_prod}, the decay of  $X$ injects energy into both the SM and dark sectors, alongside a graviton. To study the evolution of the graviton's energy density, we need to isolate the portion of energy transferred to the graviton.
Now, the 3-body decay rate of $X$ can be decomposed as~\cite{Barman:2023ymn, Kanemura:2023pnv}
\begin{equation}
	\Gamma^{1\rightarrow 3}=\int\frac{d\Gamma^{1\rightarrow 3}}{dE_{\text{gw}}}\frac{m_X-E_{\text{gw}}}{m_X}dE_{\text{gw}}+\int\frac{d\Gamma^{1\rightarrow 3}}{dE_{\text{gw}}}\frac{E_{\text{gw}}}{m_X}dE_{\text{gw}},
\end{equation} 
The second term on the \textit{r.h.s.} of this equation quantifies the energy injected into the graviton from the decay of $X$.
Notably, the differential decay rate of the graviton is determined by two parameters, $y$ and $m_X$ (see Eq.~(\ref{diff_decay_width})), which are solely fixed by the DM relic density constraint. Although the inclusion of 3-body decay to the freeze-in production is practically negligible, the GW spectrum and freeze-in are intrinsically connected, as both depend entirely on the same set of parameters. The production and evolution of the energy density of GWs in the form of graviton radiation can be described by the following Boltzmann equation:
\begin{equation}
	\frac{d\rho_{\text{gw}}}{dt}+4H \rho_{\text{gw}}=\bigg[\int\frac{d\Gamma^{1\rightarrow 3}}{dE_{\text{gw}}}\frac{E_{\text{gw}}}{m_X}dE_{\text{gw}}\bigg] \rho_{X}^{\text{eq}},
	\label{GW_evol}
\end{equation}
Here, $\rho_{X}^{\text{eq}}(=n_X^{\text{eq}}E_X)$ refers to the equilibrium energy density and $E_X(=\sqrt{m_X^2+9T^2})$ is the energy of $X$ in thermal bath~\cite{2006hep.ph....8347S}. Conventionally, it is more useful to express the evolution equation for the differential GW energy density as various GW detectors target different frequency ranges. With this in mind, and changing the variable from time to the dimensionless quantity $z$, Eq.~(\ref{GW_evol}) can be re-expressed as:
\begin{equation}
 	z\frac{d}{dz}\bigg(\frac{d\rho_{\text{gw}}}{dE_{\text{gw}}}\bigg)+4 \frac{d\rho_{\text{gw}}}{dE_{\text{gw}}}=\frac{d\Gamma^{1\rightarrow 3}}{dE_{\text{gw}}}\frac{E_{\text{gw}}}{H}\sqrt{1+\frac{9}{z^2}} n_{X}^{\text{eq}},
 	\label{GW_evol_z}
 \end{equation}
Here, GW production occurs via the 3-body decay of $X$ and saturates at point $z_s$ once the equilibrium number density of $X$ is significantly diluted, mirroring the freeze-in DM scenario discussed earlier. Once GW generation ceases, the produced gravitons propagate freely until today, forming a stochastic GW background. Importantly, the energy of the produced gravitons redshifts due to the universe's cosmic expansion. One can obtain the quantity $\frac{d\rho_{\text{gw}}}{dE_{\text{gw}}}$ at the saturation point $z_s$ by numerically solving Eq.~(\ref{GW_evol_z}), and its present-day value by considering the course of redshift.
The present-day GW energy density can be evaluated using the expression:
\begin{equation}
 	\Omega_{\text{gw}}h^2=\bigg[\frac{h^2}{\rho_c}E_{\text{gw}}\frac{d\rho_{\text{gw}}}{dE_{\text{gw}}}\bigg]_0=h^2\bigg(\frac{\Omega_\gamma^0}{\rho_R^s}\bigg)E_{\text{gw}}^s\bigg[\frac{d\rho_{\text{gw}}}{dE_{\text{gw}}}\bigg]_s,
 \end{equation}
 where $\Omega_\gamma^0(=5.46\times 10^{-5})$ represents the current relic density of photon and $E_{\text{gw}}^s=E_{\text{gw}}^0(a_0/a_s)=E_{\text{gw}}^0(\rho_R^s/\rho_R^0)^{1/4}$ refers to the energy density of mono-graviton at $z_s$ connected with the present energy of the same $E_{\text{gw}}^0=2\pi f$.
It is important to note that the upper limits on the emitted graviton energy  ($E_{\text{gw}}^{\text{max}}=m_X/2$) impose an upper cut-off on the produced GW frequency. 
Thus the maximum frequency that solely depends on $z_s$, is given by:
\begin{equation}
	f_\text{max}=\frac{m_X}{4\pi}\bigg(\frac{30\rho_R^0}{\pi^2g_*(z_s)}\bigg)^{1/4}\frac{z_s}{m_X}\simeq 3.19\times 10^{11}~\text{Hz}.
\end{equation}
where, $\rho_R^0(=(2.14\times 10^{-13}~\text{GeV})^4)$ represents the present day radiation energy density, the effective number of relativistic degrees of freedom $g_*(z_s)=g_*=106.75$ derived at saturation point $z_s=30$. 
{\sl After reaching a peak frequency of approximately $10^{11}$ Hz in our scenario, the relic of GW drops sharply to zero. Notably, this peak frequency is independent of the model parameters ($y, m_X$) (see, Supplemental Material for further clarification)}, a characteristic hallmark compared to GW spectra that bear the imprints of inflationary reheating~\cite{Nakayama:2018ptw, Huang:2019lgd, Barman:2023ymn, Barman:2023rpg, Bernal:2023wus, Tokareva:2023mrt, Bernal:2025lxp}, heavy particle decay~\cite{Kanemura:2023pnv, Choi:2024acs} or, leptogenesis~\cite{Datta:2024tne}.

For demonstrating the GW spectrum, we selected benchmark points (BPs) satisfying the relic density constraint for a fixed DM mass of $m_\chi=12$ keV. Our findings are displayed in Figure~\ref{fig:GW}, with three BPs satisfying the relic density constraint. The BPs are BP1 ($y=7.133\times10^{-4}$, $m_X=10^{12}$), BP2 ($y=1.243\times10^{-2}$, $m_X=3\times 10^{14}$) and  BP3 ($y=5.22\times10^{-2}$, $m_X=5\times 10^{15}$) Considering a maximum reheating temperature is around $5\times10^{15}$ GeV, one assumes $m_X$ can be as large as that scale.
Such a choice of dark matter mass, along with the mediator mass, ranges from $10^{12}$ to $10^{15}$ GeV, looks hierarchical, but a similar level of mass hierarchy already exists in the SM fermionic sector, or BSM scenarios like Type-I seesaw.

The future sensitivity ranges of space-based laser interferometer experiments and ground based experiments, such as, LISA~\cite{amaro2017laser}, UDECIGO~\cite{Seto:2001qf}, LIGO~\cite{KAGRA:2013rdx} Einstein Telescope (ET)~\cite{Hild:2008ng} and  Cosmic Explorer (CE)~\cite{LIGOScientific:2016wof}, operating in the intermediate frequency range ($10^{-6}$-$10^4$ Hz), are indicated by the color shaded regions, respectively. Constraints on strain were transformed to GW energy density using $\Omega_{\text{gw}}h^2 = f^2h_c^2/(1.26\times10^{-18})$ where $h_c$ refers to strain~\cite{Maggiore:1999vm}. Additionally, proposed experiments probing higher frequency ranges, spanning $10^{4}$-$10^9$ Hz using resonant cavity techniques~\cite{Herman:2020wao, Herman:2022fau}, are embedded (shown in red). 
Furthermore, the energy density generated by GWs acts as radiation before Big Bang Nucleosynthesis (BBN) and contributes to the effective neutrino count ($N_{\text{eff}}$). Consequently, the lower limit imposed by the combined analysis of BBN and Cosmic Microwave Background (CMB) data~\cite{Yeh:2022heq} on $\Delta N_{\text{eff}}(\le 0.14)$) rules out the indicated grey hatched region in Figure~\ref{fig:GW}. The future COrE/Euclid~\cite{armitage2011core, laureijs2011euclid} experiment would impose a more stringent constraint on $\Delta N_{\text{eff}}(\le 0.017)$, which would further refine the limit as shown by the extended grey-shaded area. We find that the GW energy density for $m_X=10^{12}$ GeV (BP-1) falls significantly below the sensitivity regime of resonant cavity experiments. In contrast, the GW spectra corresponding to BP-2 and BP-3 are marginally within and well inside, respectively, the GW probing range of the aforementioned experiments. We also note that for heavier DM, the GW spectrum shown in Figure~\ref{fig:GW} will shift downward, as a smaller coupling is required to satisfy the observed DM relic density.
The Freeze-in GW (FI-GW) signal would coexist with the gravitational wave produced by SM scattering in the bath, which is also known as the cosmic gravitational wave background (CGWB). Although revisiting such a study falls beyond the scope of this letter, several works with a detailed framework already exist in the literature~\cite{Ghiglieri:2015nfa, Ghiglieri:2020mhm, Ringwald:2020ist, Drewes:2023oxg}. In Figure~\ref{fig:GW}, the CGWB contribution is indicated through a grey line~\cite{Drewes:2023oxg}. Precise knowledge of the spectral slope, peak frequency, and endpoint of both spectra can be instrumental for the probe. Evidently, they provide very distinct characteristics. At lower frequencies, CGWB falls much faster, furnishing FI-GW excess over an extensive frequency range, whereas the same spectrum falls slowly after reaching its peak.  
The characteristic peak frequency of FI-GW is independent of $m_X$, which can be further enhanced when modification occurs in background cosmology~\cite{Konar:2025gvh}.

In Figure~\ref{fig:GW}, we present the GW spectrum for our benchmark values in black lines and the cosmic gravitational wave background (CGWB) generated by the SM bath, as shown in an orange line. Here, it is clear that our GW spectrum shows a very distinct pattern. For the lower frequency regime, the frequency dependence is much steeper for CGWB compared to our spectrum, and the CGWB spectrum falls slowly with frequency after reaching its peak frequency, unlike our case. Hence the freeze-in GW signal shows a clear excess over the CGWB spectrum in the low-frequency range. Moreover, freeze-in GW spectra exhibit a particular spectral slope, peak frequency and end point at higher frequency distinct from CGWB. 

\begin{figure}[t]
	\centering
	\includegraphics[width=8cm, height=6cm]{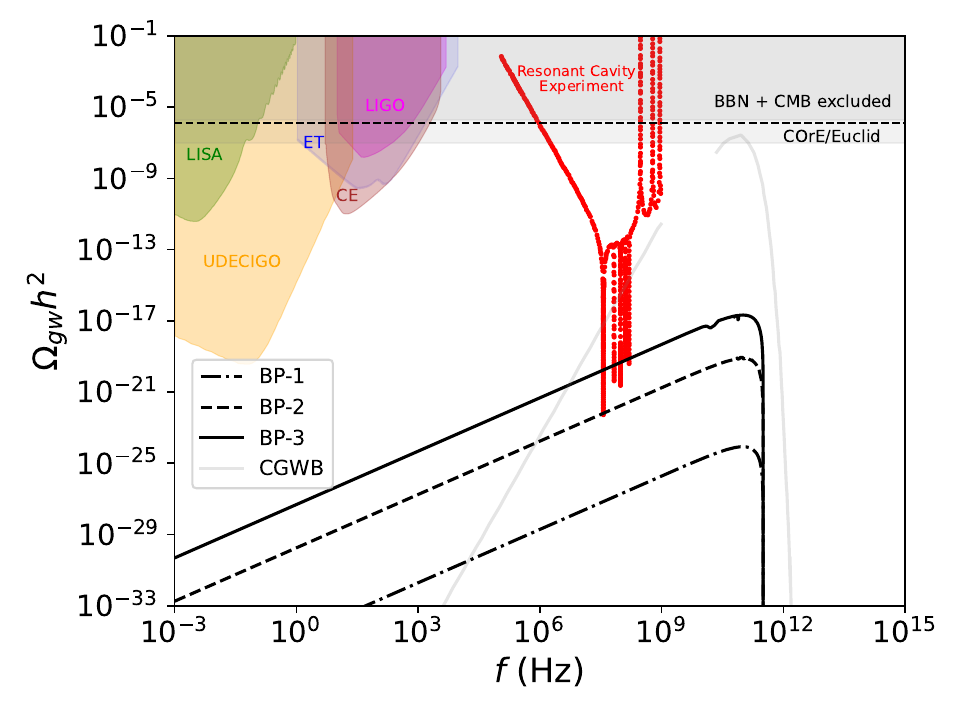}
	\caption{GW spectrum as a result of graviton \textit{bremsstrahlung} during DM freeze-in production for three BPs and CGWB spectrum.}
	\label{fig:GW}
\end{figure}

While Figure~\ref{fig:GW} indicates that resonant cavity experiments can test mediator masses for a range between $3.5\times10^{14}$ GeV and $5\times 10^{15}$ GeV for a fixed DM mass of $12$ keV, this observable mass range for the mediator is, in fact, dependent on the DM mass.
Figure~\ref{fig:summary} illustrates the range of mediator masses detectable by resonant cavity experiments as a function of DM mass. 
As the DM mass increases, a smaller coupling is required to satisfy the DM relic density constraint. This, in turn, lowers the GW relic density for a fixed mediator mass. Consequently, the probable range of heavy mediator masses that can be probed shrinks as the DM mass increases, as is evident from Figure~\ref{fig:summary}. It is crucial to emphasize that the generated GWs cannot be detected by the aforementioned experiments for DM masses exceeding 2.24 MeV, which corresponds to the right corner of the triangular region.
The grey hatched region is excluded by Lyman-$\alpha$ observations, which set a lower limit of 5.3 keV on DM mass.
The dashed vertical line corresponds to a 12 keV DM mass, demonstrating the lowest and highest points of probing triangular region, which is also represented by BP-2 and BP-3 from Figure~\ref{fig:GW}, respectively. 

To differentiate the origins of GW spectra, it is essential to measure these spectra and their associated peak frequencies across various spectral ranges that are sensitive to one or more experiments.
Currently, the existing LIGO~\cite{KAGRA:2013rdx} and proposed UDECIGO\cite{Seto:2001qf} detectors cannot detect the tail part of the GW spectra; however, the sensitivity of laser interferometers can be improved by harnessing the full capabilities of quantum sensing techniques in the future~\cite{TitoDAgnolo:2024res}. This improvement would be necessary to reach the low-frequency tail of the heavy particle bremsstrahlung GW spectrum. The stochastic GW background will partially convert into photons owing to its interaction with galactic and intergalactic magnetic fields. Interestingly, the future experiments designed to measure spectral distortions in the CMB will be sensitive to the excess photons resulting from this graviton-photon conversion, thereby allowing for the indirect detection of high-frequency gravitational waves. Voyage 2050~\cite{He:2023xoh} includes a CMB spectral distortion survey with a few nano kelvins accuracy. Observing the high-frequency peak of the bremsstrahlung GW spectrum through CMB spectral distortion surveys will require sensitivity improvements that go beyond the capabilities of Voyage 2050 or other high frequency GW experiments~\cite{Herman:2020wao, Herman:2022fau, Domcke:2022rgu}. Overall, all detectors need enhancements in sensitivity to effectively detect the GW spectra, which is required to distinguish the GW spectra bearing the footprints of different origins.

\begin{figure}[t]
	\centering
	\includegraphics[width=8cm, height=6cm]{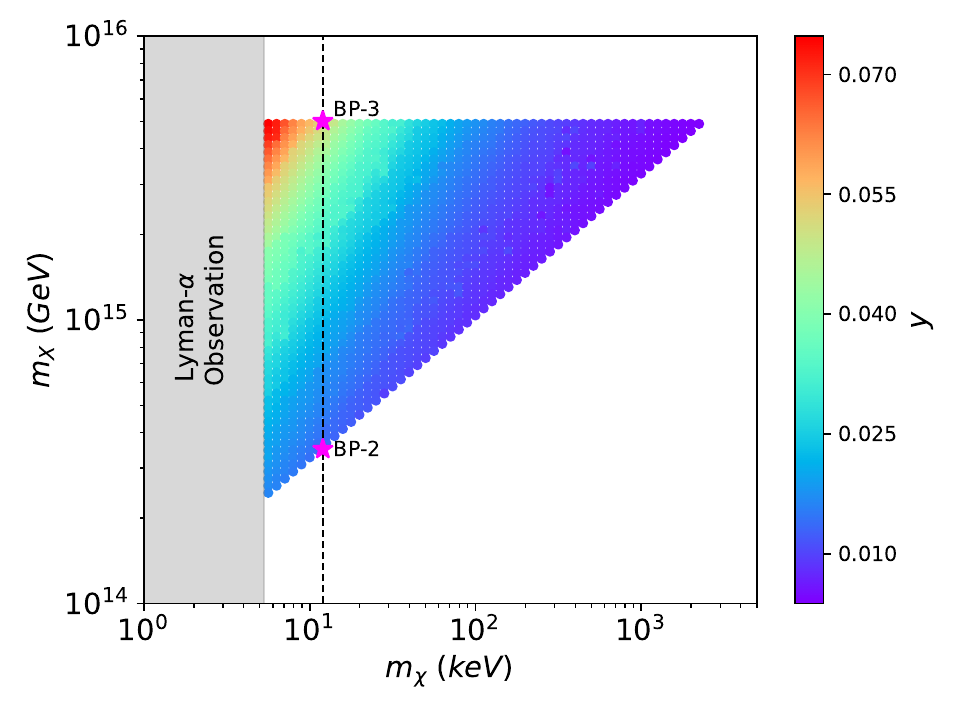}
	\caption{Range of masses probed by the planned resonant cavity experiment and constant with freeze-in relic density.} 
	\label{fig:summary}
\end{figure}

\textit{Summary and outlook}--Our study demonstrates the feasibility of probing freeze-in dark matter through gravitational waves generated as graviton radiation during DM production through the decay of heavy particles. While this has been shown for models with scalar DM and vector-like fermionic heavy particles, this approach is potentially applicable to any models featuring the interaction of the form $yX\xi\chi$. For instance, the specific results may differ by a numerical factor for fermionic dark matter.

Several high-frequency GW experiments have been proposed, with resonant cavity experiments currently showing promise in detecting GWs that bear the distinct footprints of freeze-in DM. 
This study demonstrates that GWs generated from the decay of a heavy mediator, with masses ranging from $2.5\times10^{14}$ to  $5\times10^{15}$ GeV, fall within the projected sensitivity of this experiment for a DM mass of approximately 5.4 keV, provided sub-percent accuracy can be achieved. Additionally, such a GW signal exhibits a distinct low-frequency tail in the GW spectrum. For more details detection prospects of the GW spectrum, see Supplemental Material. 
The detectable range narrowed significantly, limiting detection to a mediator mass at $m_X=5\times10^{15}$ for $m_\chi=2.24$ MeV. Beyond this point, the produced GWs are too faint to be probed by this experiment, irrespective of the mediator mass.
Interestingly, GWs can also be emitted during freeze-in DM production via scattering processes; however, such calculations are more involved than those presented here. Whether such a scenario would yield detectable signatures requires careful investigation, which we leave for future study.
Theoretically, our conclusion about the GW spectrum remains valid even for a relatively lighter mediator mass (for example, 1 TeV) and in the context of scattering production. However, this comes at the cost of a significantly smaller GW relic density.

Finally, the study of gravitational wave spectra associated with freeze-in DM is opening up novel avenues to explore this elusive freeze-in DM paradigm, which is incredibly challenging to reach through any ongoing or upcoming direct and indirect DM detection experiments, while collider experiments can only access a narrow region of parameter space.

\vspace{0.2cm}
\textit{Acknowledgment}---SS is supported by NPDF grant PDF/2023/002076 from the Science and Engineering Research Board (SERB), Government of India. The work of PK at Physical Research Laboratory (PRL) was supported by Department of Space, Government of India.

\vspace{0.2cm}
\textit{Data availability}---The data that support the findings of this article are not publicly available. The data are available from the authors upon reasonable request.
\bibliographystyle{apsrev4-2}
\bibliography{ref}

\begin{thebibliography}{61}%
\makeatletter
\providecommand \@ifxundefined [1]{%
 \@ifx{#1\undefined}
}%
\providecommand \@ifnum [1]{%
 \ifnum #1\expandafter \@firstoftwo
 \else \expandafter \@secondoftwo
 \fi
}%
\providecommand \@ifx [1]{%
 \ifx #1\expandafter \@firstoftwo
 \else \expandafter \@secondoftwo
 \fi
}%
\providecommand \natexlab [1]{#1}%
\providecommand \enquote  [1]{``#1''}%
\providecommand \bibnamefont  [1]{#1}%
\providecommand \bibfnamefont [1]{#1}%
\providecommand \citenamefont [1]{#1}%
\providecommand \href@noop [0]{\@secondoftwo}%
\providecommand \href [0]{\begingroup \@sanitize@url \@href}%
\providecommand \@href[1]{\@@startlink{#1}\@@href}%
\providecommand \@@href[1]{\endgroup#1\@@endlink}%
\providecommand \@sanitize@url [0]{\catcode `\\12\catcode `\$12\catcode
  `\&12\catcode `\#12\catcode `\^12\catcode `\_12\catcode `\%12\relax}%
\providecommand \@@startlink[1]{}%
\providecommand \@@endlink[0]{}%
\providecommand \url  [0]{\begingroup\@sanitize@url \@url }%
\providecommand \@url [1]{\endgroup\@href {#1}{\urlprefix }}%
\providecommand \urlprefix  [0]{URL }%
\providecommand \Eprint [0]{\href }%
\providecommand \doibase [0]{https://doi.org/}%
\providecommand \selectlanguage [0]{\@gobble}%
\providecommand \bibinfo  [0]{\@secondoftwo}%
\providecommand \bibfield  [0]{\@secondoftwo}%
\providecommand \translation [1]{[#1]}%
\providecommand \BibitemOpen [0]{}%
\providecommand \bibitemStop [0]{}%
\providecommand \bibitemNoStop [0]{.\EOS\space}%
\providecommand \EOS [0]{\spacefactor3000\relax}%
\providecommand \BibitemShut  [1]{\csname bibitem#1\endcsname}%
\let\auto@bib@innerbib\@empty
\bibitem [{\citenamefont {Arcadi}\ \emph {et~al.}(2018)\citenamefont {Arcadi},
  \citenamefont {Dutra}, \citenamefont {Ghosh}, \citenamefont {Lindner},
  \citenamefont {Mambrini}, \citenamefont {Pierre}, \citenamefont {Profumo},\
  and\ \citenamefont {Queiroz}}]{Arcadi:2017kky}%
  \BibitemOpen
  \bibfield  {author} {\bibinfo {author} {\bibfnamefont {G.}~\bibnamefont
  {Arcadi}}, \bibinfo {author} {\bibfnamefont {M.}~\bibnamefont {Dutra}},
  \bibinfo {author} {\bibfnamefont {P.}~\bibnamefont {Ghosh}}, \bibinfo
  {author} {\bibfnamefont {M.}~\bibnamefont {Lindner}}, \bibinfo {author}
  {\bibfnamefont {Y.}~\bibnamefont {Mambrini}}, \bibinfo {author}
  {\bibfnamefont {M.}~\bibnamefont {Pierre}}, \bibinfo {author} {\bibfnamefont
  {S.}~\bibnamefont {Profumo}},\ and\ \bibinfo {author} {\bibfnamefont {F.~S.}\
  \bibnamefont {Queiroz}},\ }\href
  {https://doi.org/10.1140/epjc/s10052-018-5662-y} {\bibfield  {journal}
  {\bibinfo  {journal} {Eur. Phys. J. C}\ }\textbf {\bibinfo {volume} {78}},\
  \bibinfo {pages} {203} (\bibinfo {year} {2018})},\ \Eprint
  {https://arxiv.org/abs/1703.07364} {arXiv:1703.07364 [hep-ph]} \BibitemShut
  {NoStop}%
\bibitem [{\citenamefont {Roszkowski}\ \emph {et~al.}(2018)\citenamefont
  {Roszkowski}, \citenamefont {Sessolo},\ and\ \citenamefont
  {Trojanowski}}]{Roszkowski:2017nbc}%
  \BibitemOpen
  \bibfield  {author} {\bibinfo {author} {\bibfnamefont {L.}~\bibnamefont
  {Roszkowski}}, \bibinfo {author} {\bibfnamefont {E.~M.}\ \bibnamefont
  {Sessolo}},\ and\ \bibinfo {author} {\bibfnamefont {S.}~\bibnamefont
  {Trojanowski}},\ }\href {https://doi.org/10.1088/1361-6633/aab913} {\bibfield
   {journal} {\bibinfo  {journal} {Rept. Prog. Phys.}\ }\textbf {\bibinfo
  {volume} {81}},\ \bibinfo {pages} {066201} (\bibinfo {year} {2018})},\
  \Eprint {https://arxiv.org/abs/1707.06277} {arXiv:1707.06277 [hep-ph]}
  \BibitemShut {NoStop}%
\bibitem [{\citenamefont {Hall}\ \emph {et~al.}(2010)\citenamefont {Hall},
  \citenamefont {Jedamzik}, \citenamefont {March-Russell},\ and\ \citenamefont
  {West}}]{Hall:2009bx}%
  \BibitemOpen
  \bibfield  {author} {\bibinfo {author} {\bibfnamefont {L.~J.}\ \bibnamefont
  {Hall}}, \bibinfo {author} {\bibfnamefont {K.}~\bibnamefont {Jedamzik}},
  \bibinfo {author} {\bibfnamefont {J.}~\bibnamefont {March-Russell}},\ and\
  \bibinfo {author} {\bibfnamefont {S.~M.}\ \bibnamefont {West}},\ }\href
  {https://doi.org/10.1007/JHEP03(2010)080} {\bibfield  {journal} {\bibinfo
  {journal} {JHEP}\ }\textbf {\bibinfo {volume} {03}},\ \bibinfo {pages}
  {080}},\ \Eprint {https://arxiv.org/abs/0911.1120} {arXiv:0911.1120 [hep-ph]}
  \BibitemShut {NoStop}%
\bibitem [{\citenamefont {Bernal}\ \emph {et~al.}(2017)\citenamefont {Bernal},
  \citenamefont {Heikinheimo}, \citenamefont {Tenkanen}, \citenamefont
  {Tuominen},\ and\ \citenamefont {Vaskonen}}]{Bernal:2017kxu}%
  \BibitemOpen
  \bibfield  {author} {\bibinfo {author} {\bibfnamefont {N.}~\bibnamefont
  {Bernal}}, \bibinfo {author} {\bibfnamefont {M.}~\bibnamefont {Heikinheimo}},
  \bibinfo {author} {\bibfnamefont {T.}~\bibnamefont {Tenkanen}}, \bibinfo
  {author} {\bibfnamefont {K.}~\bibnamefont {Tuominen}},\ and\ \bibinfo
  {author} {\bibfnamefont {V.}~\bibnamefont {Vaskonen}},\ }\href
  {https://doi.org/10.1142/S0217751X1730023X} {\bibfield  {journal} {\bibinfo
  {journal} {Int. J. Mod. Phys. A}\ }\textbf {\bibinfo {volume} {32}},\
  \bibinfo {pages} {1730023} (\bibinfo {year} {2017})},\ \Eprint
  {https://arxiv.org/abs/1706.07442} {arXiv:1706.07442 [hep-ph]} \BibitemShut
  {NoStop}%
\bibitem [{\citenamefont {Aalbers}\ \emph {et~al.}(2023)\citenamefont {Aalbers}
  \emph {et~al.}}]{LZ:2022lsv}%
  \BibitemOpen
  \bibfield  {author} {\bibinfo {author} {\bibfnamefont {J.}~\bibnamefont
  {Aalbers}} \emph {et~al.} (\bibinfo {collaboration} {LZ}),\ }\href
  {https://doi.org/10.1103/PhysRevLett.131.041002} {\bibfield  {journal}
  {\bibinfo  {journal} {Phys. Rev. Lett.}\ }\textbf {\bibinfo {volume} {131}},\
  \bibinfo {pages} {041002} (\bibinfo {year} {2023})},\ \Eprint
  {https://arxiv.org/abs/2207.03764} {arXiv:2207.03764 [hep-ex]} \BibitemShut
  {NoStop}%
\bibitem [{\citenamefont {Bo}\ \emph {et~al.}(2025)\citenamefont {Bo} \emph
  {et~al.}}]{PandaX:2024qfu}%
  \BibitemOpen
  \bibfield  {author} {\bibinfo {author} {\bibfnamefont {Z.}~\bibnamefont {Bo}}
  \emph {et~al.} (\bibinfo {collaboration} {PandaX}),\ }\href
  {https://doi.org/10.1103/PhysRevLett.134.011805} {\bibfield  {journal}
  {\bibinfo  {journal} {Phys. Rev. Lett.}\ }\textbf {\bibinfo {volume} {134}},\
  \bibinfo {pages} {011805} (\bibinfo {year} {2025})},\ \Eprint
  {https://arxiv.org/abs/2408.00664} {arXiv:2408.00664 [hep-ex]} \BibitemShut
  {NoStop}%
\bibitem [{\citenamefont {Aprile}\ \emph {et~al.}(2025)\citenamefont {Aprile}
  \emph {et~al.}}]{XENON:2025vwd}%
  \BibitemOpen
  \bibfield  {author} {\bibinfo {author} {\bibfnamefont {E.}~\bibnamefont
  {Aprile}} \emph {et~al.} (\bibinfo {collaboration} {XENON}),\ }\href@noop {}
  {\  (\bibinfo {year} {2025})},\ \Eprint {https://arxiv.org/abs/2502.18005}
  {arXiv:2502.18005 [hep-ex]} \BibitemShut {NoStop}%
\bibitem [{\citenamefont {Ackermann}\ \emph {et~al.}(2015)\citenamefont
  {Ackermann} \emph {et~al.}}]{Fermi-LAT:2015att}%
  \BibitemOpen
  \bibfield  {author} {\bibinfo {author} {\bibfnamefont {M.}~\bibnamefont
  {Ackermann}} \emph {et~al.} (\bibinfo {collaboration} {Fermi-LAT}),\ }\href
  {https://doi.org/10.1103/PhysRevLett.115.231301} {\bibfield  {journal}
  {\bibinfo  {journal} {Phys. Rev. Lett.}\ }\textbf {\bibinfo {volume} {115}},\
  \bibinfo {pages} {231301} (\bibinfo {year} {2015})},\ \Eprint
  {https://arxiv.org/abs/1503.02641} {arXiv:1503.02641 [astro-ph.HE]}
  \BibitemShut {NoStop}%
\bibitem [{\citenamefont {Albert}\ \emph {et~al.}(2017)\citenamefont {Albert}
  \emph {et~al.}}]{Fermi-LAT:2016uux}%
  \BibitemOpen
  \bibfield  {author} {\bibinfo {author} {\bibfnamefont {A.}~\bibnamefont
  {Albert}} \emph {et~al.} (\bibinfo {collaboration} {Fermi-LAT, DES}),\ }\href
  {https://doi.org/10.3847/1538-4357/834/2/110} {\bibfield  {journal} {\bibinfo
   {journal} {Astrophys. J.}\ }\textbf {\bibinfo {volume} {834}},\ \bibinfo
  {pages} {110} (\bibinfo {year} {2017})},\ \Eprint
  {https://arxiv.org/abs/1611.03184} {arXiv:1611.03184 [astro-ph.HE]}
  \BibitemShut {NoStop}%
\bibitem [{\citenamefont {Aaboud}\ \emph {et~al.}(2018)\citenamefont {Aaboud}
  \emph {et~al.}}]{ATLAS:2017tny}%
  \BibitemOpen
  \bibfield  {author} {\bibinfo {author} {\bibfnamefont {M.}~\bibnamefont
  {Aaboud}} \emph {et~al.} (\bibinfo {collaboration} {ATLAS}),\ }\href
  {https://doi.org/10.1103/PhysRevD.97.052012} {\bibfield  {journal} {\bibinfo
  {journal} {Phys. Rev. D}\ }\textbf {\bibinfo {volume} {97}},\ \bibinfo
  {pages} {052012} (\bibinfo {year} {2018})},\ \Eprint
  {https://arxiv.org/abs/1710.04901} {arXiv:1710.04901 [hep-ex]} \BibitemShut
  {NoStop}%
\bibitem [{\citenamefont {Curtin}\ \emph {et~al.}(2019)\citenamefont {Curtin}
  \emph {et~al.}}]{Curtin:2018mvb}%
  \BibitemOpen
  \bibfield  {author} {\bibinfo {author} {\bibfnamefont {D.}~\bibnamefont
  {Curtin}} \emph {et~al.},\ }\href {https://doi.org/10.1088/1361-6633/ab28d6}
  {\bibfield  {journal} {\bibinfo  {journal} {Rept. Prog. Phys.}\ }\textbf
  {\bibinfo {volume} {82}},\ \bibinfo {pages} {116201} (\bibinfo {year}
  {2019})},\ \Eprint {https://arxiv.org/abs/1806.07396} {arXiv:1806.07396
  [hep-ph]} \BibitemShut {NoStop}%
\bibitem [{\citenamefont {B\'elanger}\ \emph {et~al.}(2019)\citenamefont
  {B\'elanger} \emph {et~al.}}]{Belanger:2018sti}%
  \BibitemOpen
  \bibfield  {author} {\bibinfo {author} {\bibfnamefont {G.}~\bibnamefont
  {B\'elanger}} \emph {et~al.},\ }\href
  {https://doi.org/10.1007/JHEP02(2019)186} {\bibfield  {journal} {\bibinfo
  {journal} {JHEP}\ }\textbf {\bibinfo {volume} {02}},\ \bibinfo {pages}
  {186}},\ \Eprint {https://arxiv.org/abs/1811.05478} {arXiv:1811.05478
  [hep-ph]} \BibitemShut {NoStop}%
\bibitem [{\citenamefont {Garny}\ \emph {et~al.}(2018)\citenamefont {Garny},
  \citenamefont {Heisig}, \citenamefont {Hufnagel},\ and\ \citenamefont
  {L\"ulf}}]{Garny:2018icg}%
  \BibitemOpen
  \bibfield  {author} {\bibinfo {author} {\bibfnamefont {M.}~\bibnamefont
  {Garny}}, \bibinfo {author} {\bibfnamefont {J.}~\bibnamefont {Heisig}},
  \bibinfo {author} {\bibfnamefont {M.}~\bibnamefont {Hufnagel}},\ and\
  \bibinfo {author} {\bibfnamefont {B.}~\bibnamefont {L\"ulf}},\ }\href
  {https://doi.org/10.1103/PhysRevD.97.075002} {\bibfield  {journal} {\bibinfo
  {journal} {Phys. Rev. D}\ }\textbf {\bibinfo {volume} {97}},\ \bibinfo
  {pages} {075002} (\bibinfo {year} {2018})},\ \Eprint
  {https://arxiv.org/abs/1802.00814} {arXiv:1802.00814 [hep-ph]} \BibitemShut
  {NoStop}%
\bibitem [{\citenamefont {D'Eramo}\ \emph {et~al.}(2018)\citenamefont
  {D'Eramo}, \citenamefont {Fernandez},\ and\ \citenamefont
  {Profumo}}]{DEramo:2017ecx}%
  \BibitemOpen
  \bibfield  {author} {\bibinfo {author} {\bibfnamefont {F.}~\bibnamefont
  {D'Eramo}}, \bibinfo {author} {\bibfnamefont {N.}~\bibnamefont {Fernandez}},\
  and\ \bibinfo {author} {\bibfnamefont {S.}~\bibnamefont {Profumo}},\ }\href
  {https://doi.org/10.1088/1475-7516/2018/02/046} {\bibfield  {journal}
  {\bibinfo  {journal} {JCAP}\ }\textbf {\bibinfo {volume} {02}},\ \bibinfo
  {pages} {046}},\ \Eprint {https://arxiv.org/abs/1712.07453} {arXiv:1712.07453
  [hep-ph]} \BibitemShut {NoStop}%
\bibitem [{\citenamefont {Ghosh}\ \emph {et~al.}(2024)\citenamefont {Ghosh},
  \citenamefont {Konar},\ and\ \citenamefont {Show}}]{Ghosh:2024nkj}%
  \BibitemOpen
  \bibfield  {author} {\bibinfo {author} {\bibfnamefont {A.}~\bibnamefont
  {Ghosh}}, \bibinfo {author} {\bibfnamefont {P.}~\bibnamefont {Konar}},\ and\
  \bibinfo {author} {\bibfnamefont {S.}~\bibnamefont {Show}},\ }\href@noop {}
  {\  (\bibinfo {year} {2024})},\ \Eprint {https://arxiv.org/abs/2411.09464}
  {arXiv:2411.09464 [hep-ph]} \BibitemShut {NoStop}%
\bibitem [{\citenamefont {Becker}\ \emph {et~al.}(2024)\citenamefont {Becker},
  \citenamefont {Copello}, \citenamefont {Harz}, \citenamefont {Lang},\ and\
  \citenamefont {Xu}}]{Becker:2023tvd}%
  \BibitemOpen
  \bibfield  {author} {\bibinfo {author} {\bibfnamefont {M.}~\bibnamefont
  {Becker}}, \bibinfo {author} {\bibfnamefont {E.}~\bibnamefont {Copello}},
  \bibinfo {author} {\bibfnamefont {J.}~\bibnamefont {Harz}}, \bibinfo {author}
  {\bibfnamefont {J.}~\bibnamefont {Lang}},\ and\ \bibinfo {author}
  {\bibfnamefont {Y.}~\bibnamefont {Xu}},\ }\href
  {https://doi.org/10.1088/1475-7516/2024/01/053} {\bibfield  {journal}
  {\bibinfo  {journal} {JCAP}\ }\textbf {\bibinfo {volume} {01}},\ \bibinfo
  {pages} {053}},\ \Eprint {https://arxiv.org/abs/2306.17238} {arXiv:2306.17238
  [hep-ph]} \BibitemShut {NoStop}%
\bibitem [{\citenamefont {Chakraborti}\ \emph {et~al.}(2020)\citenamefont
  {Chakraborti}, \citenamefont {Martin},\ and\ \citenamefont
  {Poulose}}]{Chakraborti:2019ohe}%
  \BibitemOpen
  \bibfield  {author} {\bibinfo {author} {\bibfnamefont {S.}~\bibnamefont
  {Chakraborti}}, \bibinfo {author} {\bibfnamefont {V.}~\bibnamefont
  {Martin}},\ and\ \bibinfo {author} {\bibfnamefont {P.}~\bibnamefont
  {Poulose}},\ }\href {https://doi.org/10.1088/1475-7516/2020/03/057}
  {\bibfield  {journal} {\bibinfo  {journal} {JCAP}\ }\textbf {\bibinfo
  {volume} {03}}\bibfield  {number} {\bibinfo  {number} { (03)},\ \bibinfo
  {pages} {057}},\ }\Eprint {https://arxiv.org/abs/1904.09945}
  {arXiv:1904.09945 [hep-ph]} \BibitemShut {NoStop}%
\bibitem [{\citenamefont {No}\ \emph {et~al.}(2020)\citenamefont {No},
  \citenamefont {Tunney},\ and\ \citenamefont {Zaldivar}}]{No:2019gvl}%
  \BibitemOpen
  \bibfield  {author} {\bibinfo {author} {\bibfnamefont {J.~M.}\ \bibnamefont
  {No}}, \bibinfo {author} {\bibfnamefont {P.}~\bibnamefont {Tunney}},\ and\
  \bibinfo {author} {\bibfnamefont {B.}~\bibnamefont {Zaldivar}},\ }\href
  {https://doi.org/10.1007/JHEP03(2020)022} {\bibfield  {journal} {\bibinfo
  {journal} {JHEP}\ }\textbf {\bibinfo {volume} {03}},\ \bibinfo {pages}
  {022}},\ \Eprint {https://arxiv.org/abs/1908.11387} {arXiv:1908.11387
  [hep-ph]} \BibitemShut {NoStop}%
\bibitem [{\citenamefont {Ghosh}\ \emph {et~al.}(2022)\citenamefont {Ghosh},
  \citenamefont {Konar}, \citenamefont {Saha},\ and\ \citenamefont
  {Show}}]{Ghosh:2021wrk}%
  \BibitemOpen
  \bibfield  {author} {\bibinfo {author} {\bibfnamefont {P.}~\bibnamefont
  {Ghosh}}, \bibinfo {author} {\bibfnamefont {P.}~\bibnamefont {Konar}},
  \bibinfo {author} {\bibfnamefont {A.~K.}\ \bibnamefont {Saha}},\ and\
  \bibinfo {author} {\bibfnamefont {S.}~\bibnamefont {Show}},\ }\href
  {https://doi.org/10.1088/1475-7516/2022/10/017} {\bibfield  {journal}
  {\bibinfo  {journal} {JCAP}\ }\textbf {\bibinfo {volume} {10}},\ \bibinfo
  {pages} {017}},\ \Eprint {https://arxiv.org/abs/2112.09057} {arXiv:2112.09057
  [hep-ph]} \BibitemShut {NoStop}%
\bibitem [{\citenamefont {Das}\ \emph {et~al.}(2023)\citenamefont {Das},
  \citenamefont {Konar}, \citenamefont {Kundu},\ and\ \citenamefont
  {Show}}]{Das:2023owa}%
  \BibitemOpen
  \bibfield  {author} {\bibinfo {author} {\bibfnamefont {P.~K.}\ \bibnamefont
  {Das}}, \bibinfo {author} {\bibfnamefont {P.}~\bibnamefont {Konar}}, \bibinfo
  {author} {\bibfnamefont {S.}~\bibnamefont {Kundu}},\ and\ \bibinfo {author}
  {\bibfnamefont {S.}~\bibnamefont {Show}},\ }\href
  {https://doi.org/10.1007/JHEP06(2023)198} {\bibfield  {journal} {\bibinfo
  {journal} {JHEP}\ }\textbf {\bibinfo {volume} {06}},\ \bibinfo {pages}
  {198}},\ \Eprint {https://arxiv.org/abs/2301.02514} {arXiv:2301.02514
  [hep-ph]} \BibitemShut {NoStop}%
\bibitem [{\citenamefont {Calibbi}\ \emph {et~al.}(2015)\citenamefont
  {Calibbi}, \citenamefont {Mariotti},\ and\ \citenamefont
  {Tziveloglou}}]{Calibbi:2015nha}%
  \BibitemOpen
  \bibfield  {author} {\bibinfo {author} {\bibfnamefont {L.}~\bibnamefont
  {Calibbi}}, \bibinfo {author} {\bibfnamefont {A.}~\bibnamefont {Mariotti}},\
  and\ \bibinfo {author} {\bibfnamefont {P.}~\bibnamefont {Tziveloglou}},\
  }\href {https://doi.org/10.1007/JHEP10(2015)116} {\bibfield  {journal}
  {\bibinfo  {journal} {JHEP}\ }\textbf {\bibinfo {volume} {10}},\ \bibinfo
  {pages} {116}},\ \Eprint {https://arxiv.org/abs/1505.03867} {arXiv:1505.03867
  [hep-ph]} \BibitemShut {NoStop}%
\bibitem [{\citenamefont {Calibbi}\ \emph {et~al.}(2018)\citenamefont
  {Calibbi}, \citenamefont {Lopez-Honorez}, \citenamefont {Lowette},\ and\
  \citenamefont {Mariotti}}]{Calibbi:2018fqf}%
  \BibitemOpen
  \bibfield  {author} {\bibinfo {author} {\bibfnamefont {L.}~\bibnamefont
  {Calibbi}}, \bibinfo {author} {\bibfnamefont {L.}~\bibnamefont
  {Lopez-Honorez}}, \bibinfo {author} {\bibfnamefont {S.}~\bibnamefont
  {Lowette}},\ and\ \bibinfo {author} {\bibfnamefont {A.}~\bibnamefont
  {Mariotti}},\ }\href {https://doi.org/10.1007/JHEP09(2018)037} {\bibfield
  {journal} {\bibinfo  {journal} {JHEP}\ }\textbf {\bibinfo {volume} {09}},\
  \bibinfo {pages} {037}},\ \Eprint {https://arxiv.org/abs/1805.04423}
  {arXiv:1805.04423 [hep-ph]} \BibitemShut {NoStop}%
\bibitem [{\citenamefont {Ir\v{s}i\v{c}}\ \emph {et~al.}(2017)\citenamefont
  {Ir\v{s}i\v{c}} \emph {et~al.}}]{Irsic:2017ixq}%
  \BibitemOpen
  \bibfield  {author} {\bibinfo {author} {\bibfnamefont {V.}~\bibnamefont
  {Ir\v{s}i\v{c}}} \emph {et~al.},\ }\href
  {https://doi.org/10.1103/PhysRevD.96.023522} {\bibfield  {journal} {\bibinfo
  {journal} {Phys. Rev. D}\ }\textbf {\bibinfo {volume} {96}},\ \bibinfo
  {pages} {023522} (\bibinfo {year} {2017})},\ \Eprint
  {https://arxiv.org/abs/1702.01764} {arXiv:1702.01764 [astro-ph.CO]}
  \BibitemShut {NoStop}%
\bibitem [{\citenamefont {D'Eramo}\ and\ \citenamefont
  {Lenoci}(2021)}]{DEramo:2020gpr}%
  \BibitemOpen
  \bibfield  {author} {\bibinfo {author} {\bibfnamefont {F.}~\bibnamefont
  {D'Eramo}}\ and\ \bibinfo {author} {\bibfnamefont {A.}~\bibnamefont
  {Lenoci}},\ }\href {https://doi.org/10.1088/1475-7516/2021/10/045} {\bibfield
   {journal} {\bibinfo  {journal} {JCAP}\ }\textbf {\bibinfo {volume} {10}},\
  \bibinfo {pages} {045}},\ \Eprint {https://arxiv.org/abs/2012.01446}
  {arXiv:2012.01446 [hep-ph]} \BibitemShut {NoStop}%
\bibitem [{\citenamefont {Viel}\ \emph {et~al.}(2013)\citenamefont {Viel},
  \citenamefont {Becker}, \citenamefont {Bolton},\ and\ \citenamefont
  {Haehnelt}}]{Viel:2013fqw}%
  \BibitemOpen
  \bibfield  {author} {\bibinfo {author} {\bibfnamefont {M.}~\bibnamefont
  {Viel}}, \bibinfo {author} {\bibfnamefont {G.~D.}\ \bibnamefont {Becker}},
  \bibinfo {author} {\bibfnamefont {J.~S.}\ \bibnamefont {Bolton}},\ and\
  \bibinfo {author} {\bibfnamefont {M.~G.}\ \bibnamefont {Haehnelt}},\ }\href
  {https://doi.org/10.1103/PhysRevD.88.043502} {\bibfield  {journal} {\bibinfo
  {journal} {Phys. Rev. D}\ }\textbf {\bibinfo {volume} {88}},\ \bibinfo
  {pages} {043502} (\bibinfo {year} {2013})},\ \Eprint
  {https://arxiv.org/abs/1306.2314} {arXiv:1306.2314 [astro-ph.CO]}
  \BibitemShut {NoStop}%
\bibitem [{\citenamefont {Lovell}\ \emph {et~al.}(2012)\citenamefont {Lovell},
  \citenamefont {Eke}, \citenamefont {Frenk}, \citenamefont {Gao},
  \citenamefont {Jenkins}, \citenamefont {Theuns}, \citenamefont {Wang},
  \citenamefont {White}, \citenamefont {Boyarsky},\ and\ \citenamefont
  {Ruchayskiy}}]{Lovell_2012}%
  \BibitemOpen
  \bibfield  {author} {\bibinfo {author} {\bibfnamefont {M.~R.}\ \bibnamefont
  {Lovell}}, \bibinfo {author} {\bibfnamefont {V.}~\bibnamefont {Eke}},
  \bibinfo {author} {\bibfnamefont {C.~S.}\ \bibnamefont {Frenk}}, \bibinfo
  {author} {\bibfnamefont {L.}~\bibnamefont {Gao}}, \bibinfo {author}
  {\bibfnamefont {A.}~\bibnamefont {Jenkins}}, \bibinfo {author} {\bibfnamefont
  {T.}~\bibnamefont {Theuns}}, \bibinfo {author} {\bibfnamefont
  {J.}~\bibnamefont {Wang}}, \bibinfo {author} {\bibfnamefont {S.~D.~M.}\
  \bibnamefont {White}}, \bibinfo {author} {\bibfnamefont {A.}~\bibnamefont
  {Boyarsky}},\ and\ \bibinfo {author} {\bibfnamefont {O.}~\bibnamefont
  {Ruchayskiy}},\ }\href {https://doi.org/10.1111/j.1365-2966.2011.20200.x}
  {\bibfield  {journal} {\bibinfo  {journal} {Monthly Notices of the Royal
  Astronomical Society}\ }\textbf {\bibinfo {volume} {420}},\ \bibinfo {pages}
  {2318–2324} (\bibinfo {year} {2012})}\BibitemShut {NoStop}%
\bibitem [{\citenamefont {Bode}\ \emph {et~al.}(2001)\citenamefont {Bode},
  \citenamefont {Ostriker},\ and\ \citenamefont {Turok}}]{Bode:2000gq}%
  \BibitemOpen
  \bibfield  {author} {\bibinfo {author} {\bibfnamefont {P.}~\bibnamefont
  {Bode}}, \bibinfo {author} {\bibfnamefont {J.~P.}\ \bibnamefont {Ostriker}},\
  and\ \bibinfo {author} {\bibfnamefont {N.}~\bibnamefont {Turok}},\ }\href
  {https://doi.org/10.1086/321541} {\bibfield  {journal} {\bibinfo  {journal}
  {Astrophys. J.}\ }\textbf {\bibinfo {volume} {556}},\ \bibinfo {pages} {93}
  (\bibinfo {year} {2001})},\ \Eprint {https://arxiv.org/abs/astro-ph/0010389}
  {arXiv:astro-ph/0010389} \BibitemShut {NoStop}%
\bibitem [{\citenamefont {Choi}\ \emph {et~al.}(1995)\citenamefont {Choi},
  \citenamefont {Shim},\ and\ \citenamefont {Song}}]{Choi:1994ax}%
  \BibitemOpen
  \bibfield  {author} {\bibinfo {author} {\bibfnamefont {S.~Y.}\ \bibnamefont
  {Choi}}, \bibinfo {author} {\bibfnamefont {J.~S.}\ \bibnamefont {Shim}},\
  and\ \bibinfo {author} {\bibfnamefont {H.~S.}\ \bibnamefont {Song}},\ }\href
  {https://doi.org/10.1103/PhysRevD.51.2751} {\bibfield  {journal} {\bibinfo
  {journal} {Phys. Rev. D}\ }\textbf {\bibinfo {volume} {51}},\ \bibinfo
  {pages} {2751} (\bibinfo {year} {1995})},\ \Eprint
  {https://arxiv.org/abs/hep-th/9411092} {arXiv:hep-th/9411092} \BibitemShut
  {NoStop}%
\bibitem [{\citenamefont {Holstein}(2006)}]{Holstein:2006bh}%
  \BibitemOpen
  \bibfield  {author} {\bibinfo {author} {\bibfnamefont {B.~R.}\ \bibnamefont
  {Holstein}},\ }\href {https://doi.org/10.1119/1.2338547} {\bibfield
  {journal} {\bibinfo  {journal} {Am. J. Phys.}\ }\textbf {\bibinfo {volume}
  {74}},\ \bibinfo {pages} {1002} (\bibinfo {year} {2006})},\ \Eprint
  {https://arxiv.org/abs/gr-qc/0607045} {arXiv:gr-qc/0607045} \BibitemShut
  {NoStop}%
\bibitem [{\citenamefont {Aghanim}\ \emph {et~al.}(2020)\citenamefont {Aghanim}
  \emph {et~al.}}]{Planck:2018vyg}%
  \BibitemOpen
  \bibfield  {author} {\bibinfo {author} {\bibfnamefont {N.}~\bibnamefont
  {Aghanim}} \emph {et~al.} (\bibinfo {collaboration} {Planck}),\ }\href
  {https://doi.org/10.1051/0004-6361/201833910} {\bibfield  {journal} {\bibinfo
   {journal} {Astron. Astrophys.}\ }\textbf {\bibinfo {volume} {641}},\
  \bibinfo {pages} {A6} (\bibinfo {year} {2020})},\ \bibinfo {note} {[Erratum:
  Astron.Astrophys. 652, C4 (2021)]},\ \Eprint
  {https://arxiv.org/abs/1807.06209} {arXiv:1807.06209 [astro-ph.CO]}
  \BibitemShut {NoStop}%
\bibitem [{\citenamefont {{Strumia}}(2006)}]{2006hep.ph....8347S}%
  \BibitemOpen
  \bibfield  {author} {\bibinfo {author} {\bibfnamefont {A.}~\bibnamefont
  {{Strumia}}},\ }\href {https://doi.org/10.48550/arXiv.hep-ph/0608347}
  {\bibfield  {journal} {\bibinfo  {journal} {arXiv e-prints}\ ,\ \bibinfo
  {eid} {hep-ph/0608347}} (\bibinfo {year} {2006})},\ \Eprint
  {https://arxiv.org/abs/hep-ph/0608347} {arXiv:hep-ph/0608347 [hep-ph]}
  \BibitemShut {NoStop}%
\bibitem [{\citenamefont {{Chen}}(2007)}]{2007hep.ph....3087C}%
  \BibitemOpen
  \bibfield  {author} {\bibinfo {author} {\bibfnamefont {M.-C.}\ \bibnamefont
  {{Chen}}},\ }\href {https://doi.org/10.48550/arXiv.hep-ph/0703087} {\bibfield
   {journal} {\bibinfo  {journal} {arXiv e-prints}\ ,\ \bibinfo {eid}
  {hep-ph/0703087}} (\bibinfo {year} {2007})},\ \Eprint
  {https://arxiv.org/abs/hep-ph/0703087} {arXiv:hep-ph/0703087 [hep-ph]}
  \BibitemShut {NoStop}%
\bibitem [{\citenamefont {Barman}\ \emph
  {et~al.}(2023{\natexlab{a}})\citenamefont {Barman}, \citenamefont {Bernal},
  \citenamefont {Xu},\ and\ \citenamefont {Zapata}}]{Barman:2023ymn}%
  \BibitemOpen
  \bibfield  {author} {\bibinfo {author} {\bibfnamefont {B.}~\bibnamefont
  {Barman}}, \bibinfo {author} {\bibfnamefont {N.}~\bibnamefont {Bernal}},
  \bibinfo {author} {\bibfnamefont {Y.}~\bibnamefont {Xu}},\ and\ \bibinfo
  {author} {\bibfnamefont {O.}~\bibnamefont {Zapata}},\ }\href
  {https://doi.org/10.1088/1475-7516/2023/05/019} {\bibfield  {journal}
  {\bibinfo  {journal} {JCAP}\ }\textbf {\bibinfo {volume} {05}},\ \bibinfo
  {pages} {019}},\ \Eprint {https://arxiv.org/abs/2301.11345} {arXiv:2301.11345
  [hep-ph]} \BibitemShut {NoStop}%
\bibitem [{\citenamefont {Kanemura}\ and\ \citenamefont
  {Kaneta}(2024)}]{Kanemura:2023pnv}%
  \BibitemOpen
  \bibfield  {author} {\bibinfo {author} {\bibfnamefont {S.}~\bibnamefont
  {Kanemura}}\ and\ \bibinfo {author} {\bibfnamefont {K.}~\bibnamefont
  {Kaneta}},\ }\href {https://doi.org/10.1016/j.physletb.2024.138807}
  {\bibfield  {journal} {\bibinfo  {journal} {Phys. Lett. B}\ }\textbf
  {\bibinfo {volume} {855}},\ \bibinfo {pages} {138807} (\bibinfo {year}
  {2024})},\ \Eprint {https://arxiv.org/abs/2310.12023} {arXiv:2310.12023
  [hep-ph]} \BibitemShut {NoStop}%
\bibitem [{\citenamefont {Nakayama}\ and\ \citenamefont
  {Tang}(2019)}]{Nakayama:2018ptw}%
  \BibitemOpen
  \bibfield  {author} {\bibinfo {author} {\bibfnamefont {K.}~\bibnamefont
  {Nakayama}}\ and\ \bibinfo {author} {\bibfnamefont {Y.}~\bibnamefont
  {Tang}},\ }\href {https://doi.org/10.1016/j.physletb.2018.11.023} {\bibfield
  {journal} {\bibinfo  {journal} {Phys. Lett. B}\ }\textbf {\bibinfo {volume}
  {788}},\ \bibinfo {pages} {341} (\bibinfo {year} {2019})},\ \bibinfo {note}
  {[Erratum: Phys.Lett.B 839, 137787 (2023)]},\ \Eprint
  {https://arxiv.org/abs/1810.04975} {arXiv:1810.04975 [hep-ph]} \BibitemShut
  {NoStop}%
\bibitem [{\citenamefont {Huang}\ and\ \citenamefont
  {Yin}(2019)}]{Huang:2019lgd}%
  \BibitemOpen
  \bibfield  {author} {\bibinfo {author} {\bibfnamefont {D.}~\bibnamefont
  {Huang}}\ and\ \bibinfo {author} {\bibfnamefont {L.}~\bibnamefont {Yin}},\
  }\href {https://doi.org/10.1103/PhysRevD.100.043538} {\bibfield  {journal}
  {\bibinfo  {journal} {Phys. Rev. D}\ }\textbf {\bibinfo {volume} {100}},\
  \bibinfo {pages} {043538} (\bibinfo {year} {2019})},\ \Eprint
  {https://arxiv.org/abs/1905.08510} {arXiv:1905.08510 [hep-ph]} \BibitemShut
  {NoStop}%
\bibitem [{\citenamefont {Barman}\ \emph
  {et~al.}(2023{\natexlab{b}})\citenamefont {Barman}, \citenamefont {Bernal},
  \citenamefont {Xu},\ and\ \citenamefont {Zapata}}]{Barman:2023rpg}%
  \BibitemOpen
  \bibfield  {author} {\bibinfo {author} {\bibfnamefont {B.}~\bibnamefont
  {Barman}}, \bibinfo {author} {\bibfnamefont {N.}~\bibnamefont {Bernal}},
  \bibinfo {author} {\bibfnamefont {Y.}~\bibnamefont {Xu}},\ and\ \bibinfo
  {author} {\bibfnamefont {O.}~\bibnamefont {Zapata}},\ }\href
  {https://doi.org/10.1103/PhysRevD.108.083524} {\bibfield  {journal} {\bibinfo
   {journal} {Phys. Rev. D}\ }\textbf {\bibinfo {volume} {108}},\ \bibinfo
  {pages} {083524} (\bibinfo {year} {2023}{\natexlab{b}})},\ \Eprint
  {https://arxiv.org/abs/2305.16388} {arXiv:2305.16388 [hep-ph]} \BibitemShut
  {NoStop}%
\bibitem [{\citenamefont {Bernal}\ \emph {et~al.}(2024)\citenamefont {Bernal},
  \citenamefont {Cl\'ery}, \citenamefont {Mambrini},\ and\ \citenamefont
  {Xu}}]{Bernal:2023wus}%
  \BibitemOpen
  \bibfield  {author} {\bibinfo {author} {\bibfnamefont {N.}~\bibnamefont
  {Bernal}}, \bibinfo {author} {\bibfnamefont {S.}~\bibnamefont {Cl\'ery}},
  \bibinfo {author} {\bibfnamefont {Y.}~\bibnamefont {Mambrini}},\ and\
  \bibinfo {author} {\bibfnamefont {Y.}~\bibnamefont {Xu}},\ }\href
  {https://doi.org/10.1088/1475-7516/2024/01/065} {\bibfield  {journal}
  {\bibinfo  {journal} {JCAP}\ }\textbf {\bibinfo {volume} {01}},\ \bibinfo
  {pages} {065}},\ \Eprint {https://arxiv.org/abs/2311.12694} {arXiv:2311.12694
  [hep-ph]} \BibitemShut {NoStop}%
\bibitem [{\citenamefont {Tokareva}(2024)}]{Tokareva:2023mrt}%
  \BibitemOpen
  \bibfield  {author} {\bibinfo {author} {\bibfnamefont {A.}~\bibnamefont
  {Tokareva}},\ }\href {https://doi.org/10.1016/j.physletb.2024.138695}
  {\bibfield  {journal} {\bibinfo  {journal} {Phys. Lett. B}\ }\textbf
  {\bibinfo {volume} {853}},\ \bibinfo {pages} {138695} (\bibinfo {year}
  {2024})},\ \Eprint {https://arxiv.org/abs/2312.16691} {arXiv:2312.16691
  [hep-ph]} \BibitemShut {NoStop}%
\bibitem [{\citenamefont {Bernal}\ \emph {et~al.}(2025)\citenamefont {Bernal},
  \citenamefont {Wu}, \citenamefont {Xu},\ and\ \citenamefont
  {Xu}}]{Bernal:2025lxp}%
  \BibitemOpen
  \bibfield  {author} {\bibinfo {author} {\bibfnamefont {N.}~\bibnamefont
  {Bernal}}, \bibinfo {author} {\bibfnamefont {Q.-f.}\ \bibnamefont {Wu}},
  \bibinfo {author} {\bibfnamefont {X.-J.}\ \bibnamefont {Xu}},\ and\ \bibinfo
  {author} {\bibfnamefont {Y.}~\bibnamefont {Xu}},\ }\href@noop {} {\
  (\bibinfo {year} {2025})},\ \Eprint {https://arxiv.org/abs/2503.10756}
  {arXiv:2503.10756 [hep-ph]} \BibitemShut {NoStop}%
\bibitem [{\citenamefont {Choi}\ \emph {et~al.}(2024)\citenamefont {Choi},
  \citenamefont {Lkhagvadorj},\ and\ \citenamefont {Mahapatra}}]{Choi:2024acs}%
  \BibitemOpen
  \bibfield  {author} {\bibinfo {author} {\bibfnamefont {K.-Y.}\ \bibnamefont
  {Choi}}, \bibinfo {author} {\bibfnamefont {E.}~\bibnamefont {Lkhagvadorj}},\
  and\ \bibinfo {author} {\bibfnamefont {S.}~\bibnamefont {Mahapatra}},\ }\href
  {https://doi.org/10.1088/1475-7516/2024/07/064} {\bibfield  {journal}
  {\bibinfo  {journal} {JCAP}\ }\textbf {\bibinfo {volume} {07}},\ \bibinfo
  {pages} {064}},\ \Eprint {https://arxiv.org/abs/2403.15269} {arXiv:2403.15269
  [hep-ph]} \BibitemShut {NoStop}%
\bibitem [{\citenamefont {Datta}\ and\ \citenamefont
  {Sil}(2024)}]{Datta:2024tne}%
  \BibitemOpen
  \bibfield  {author} {\bibinfo {author} {\bibfnamefont {A.}~\bibnamefont
  {Datta}}\ and\ \bibinfo {author} {\bibfnamefont {A.}~\bibnamefont {Sil}},\
  }\href@noop {} {\  (\bibinfo {year} {2024})},\ \Eprint
  {https://arxiv.org/abs/2410.01900} {arXiv:2410.01900 [hep-ph]} \BibitemShut
  {NoStop}%
\bibitem [{\citenamefont {Amaro-Seoane}\ \emph {et~al.}(2017)\citenamefont
  {Amaro-Seoane}, \citenamefont {Audley}, \citenamefont {Babak}, \citenamefont
  {Baker}, \citenamefont {Barausse}, \citenamefont {Bender}, \citenamefont
  {Berti}, \citenamefont {Binetruy}, \citenamefont {Born}, \citenamefont
  {Bortoluzzi} \emph {et~al.}}]{amaro2017laser}%
  \BibitemOpen
  \bibfield  {author} {\bibinfo {author} {\bibfnamefont {P.}~\bibnamefont
  {Amaro-Seoane}}, \bibinfo {author} {\bibfnamefont {H.}~\bibnamefont
  {Audley}}, \bibinfo {author} {\bibfnamefont {S.}~\bibnamefont {Babak}},
  \bibinfo {author} {\bibfnamefont {J.}~\bibnamefont {Baker}}, \bibinfo
  {author} {\bibfnamefont {E.}~\bibnamefont {Barausse}}, \bibinfo {author}
  {\bibfnamefont {P.}~\bibnamefont {Bender}}, \bibinfo {author} {\bibfnamefont
  {E.}~\bibnamefont {Berti}}, \bibinfo {author} {\bibfnamefont
  {P.}~\bibnamefont {Binetruy}}, \bibinfo {author} {\bibfnamefont
  {M.}~\bibnamefont {Born}}, \bibinfo {author} {\bibfnamefont {D.}~\bibnamefont
  {Bortoluzzi}}, \emph {et~al.},\ }\href@noop {} {\bibfield  {journal}
  {\bibinfo  {journal} {arXiv preprint arXiv:1702.00786}\ } (\bibinfo {year}
  {2017})}\BibitemShut {NoStop}%
\bibitem [{\citenamefont {Seto}\ \emph {et~al.}(2001)\citenamefont {Seto},
  \citenamefont {Kawamura},\ and\ \citenamefont {Nakamura}}]{Seto:2001qf}%
  \BibitemOpen
  \bibfield  {author} {\bibinfo {author} {\bibfnamefont {N.}~\bibnamefont
  {Seto}}, \bibinfo {author} {\bibfnamefont {S.}~\bibnamefont {Kawamura}},\
  and\ \bibinfo {author} {\bibfnamefont {T.}~\bibnamefont {Nakamura}},\ }\href
  {https://doi.org/10.1103/PhysRevLett.87.221103} {\bibfield  {journal}
  {\bibinfo  {journal} {Phys. Rev. Lett.}\ }\textbf {\bibinfo {volume} {87}},\
  \bibinfo {pages} {221103} (\bibinfo {year} {2001})},\ \Eprint
  {https://arxiv.org/abs/astro-ph/0108011} {arXiv:astro-ph/0108011}
  \BibitemShut {NoStop}%
\bibitem [{\citenamefont {Abbott}\ \emph {et~al.}(2016)\citenamefont {Abbott}
  \emph {et~al.}}]{KAGRA:2013rdx}%
  \BibitemOpen
  \bibfield  {author} {\bibinfo {author} {\bibfnamefont {B.~P.}\ \bibnamefont
  {Abbott}} \emph {et~al.} (\bibinfo {collaboration} {KAGRA, LIGO Scientific,
  Virgo}),\ }\href {https://doi.org/10.1007/s41114-020-00026-9} {\bibfield
  {journal} {\bibinfo  {journal} {Living Rev. Rel.}\ }\textbf {\bibinfo
  {volume} {19}},\ \bibinfo {pages} {1} (\bibinfo {year} {2016})},\ \Eprint
  {https://arxiv.org/abs/1304.0670} {arXiv:1304.0670 [gr-qc]} \BibitemShut
  {NoStop}%
\bibitem [{\citenamefont {Hild}\ \emph {et~al.}(2008)\citenamefont {Hild},
  \citenamefont {Chelkowski},\ and\ \citenamefont {Freise}}]{Hild:2008ng}%
  \BibitemOpen
  \bibfield  {author} {\bibinfo {author} {\bibfnamefont {S.}~\bibnamefont
  {Hild}}, \bibinfo {author} {\bibfnamefont {S.}~\bibnamefont {Chelkowski}},\
  and\ \bibinfo {author} {\bibfnamefont {A.}~\bibnamefont {Freise}},\
  }\href@noop {} {\  (\bibinfo {year} {2008})},\ \Eprint
  {https://arxiv.org/abs/0810.0604} {arXiv:0810.0604 [gr-qc]} \BibitemShut
  {NoStop}%
\bibitem [{\citenamefont {Abbott}\ \emph {et~al.}(2017)\citenamefont {Abbott}
  \emph {et~al.}}]{LIGOScientific:2016wof}%
  \BibitemOpen
  \bibfield  {author} {\bibinfo {author} {\bibfnamefont {B.~P.}\ \bibnamefont
  {Abbott}} \emph {et~al.} (\bibinfo {collaboration} {LIGO Scientific}),\
  }\href {https://doi.org/10.1088/1361-6382/aa51f4} {\bibfield  {journal}
  {\bibinfo  {journal} {Class. Quant. Grav.}\ }\textbf {\bibinfo {volume}
  {34}},\ \bibinfo {pages} {044001} (\bibinfo {year} {2017})},\ \Eprint
  {https://arxiv.org/abs/1607.08697} {arXiv:1607.08697 [astro-ph.IM]}
  \BibitemShut {NoStop}%
\bibitem [{\citenamefont {Maggiore}(2000)}]{Maggiore:1999vm}%
  \BibitemOpen
  \bibfield  {author} {\bibinfo {author} {\bibfnamefont {M.}~\bibnamefont
  {Maggiore}},\ }\href {https://doi.org/10.1016/S0370-1573(99)00102-7}
  {\bibfield  {journal} {\bibinfo  {journal} {Phys. Rept.}\ }\textbf {\bibinfo
  {volume} {331}},\ \bibinfo {pages} {283} (\bibinfo {year} {2000})},\ \Eprint
  {https://arxiv.org/abs/gr-qc/9909001} {arXiv:gr-qc/9909001} \BibitemShut
  {NoStop}%
\bibitem [{\citenamefont {Herman}\ \emph {et~al.}(2021)\citenamefont {Herman},
  \citenamefont {F\"uzfa}, \citenamefont {Lehoucq},\ and\ \citenamefont
  {Clesse}}]{Herman:2020wao}%
  \BibitemOpen
  \bibfield  {author} {\bibinfo {author} {\bibfnamefont {N.}~\bibnamefont
  {Herman}}, \bibinfo {author} {\bibfnamefont {A.}~\bibnamefont {F\"uzfa}},
  \bibinfo {author} {\bibfnamefont {L.}~\bibnamefont {Lehoucq}},\ and\ \bibinfo
  {author} {\bibfnamefont {S.}~\bibnamefont {Clesse}},\ }\href
  {https://doi.org/10.1103/PhysRevD.104.023524} {\bibfield  {journal} {\bibinfo
   {journal} {Phys. Rev. D}\ }\textbf {\bibinfo {volume} {104}},\ \bibinfo
  {pages} {023524} (\bibinfo {year} {2021})},\ \Eprint
  {https://arxiv.org/abs/2012.12189} {arXiv:2012.12189 [gr-qc]} \BibitemShut
  {NoStop}%
\bibitem [{\citenamefont {Herman}\ \emph {et~al.}(2023)\citenamefont {Herman},
  \citenamefont {Lehoucq},\ and\ \citenamefont {F\'{u}zfa}}]{Herman:2022fau}%
  \BibitemOpen
  \bibfield  {author} {\bibinfo {author} {\bibfnamefont {N.}~\bibnamefont
  {Herman}}, \bibinfo {author} {\bibfnamefont {L.}~\bibnamefont {Lehoucq}},\
  and\ \bibinfo {author} {\bibfnamefont {A.}~\bibnamefont {F\'{u}zfa}},\ }\href
  {https://doi.org/10.1103/PhysRevD.108.124009} {\bibfield  {journal} {\bibinfo
   {journal} {Phys. Rev. D}\ }\textbf {\bibinfo {volume} {108}},\ \bibinfo
  {pages} {124009} (\bibinfo {year} {2023})},\ \Eprint
  {https://arxiv.org/abs/2203.15668} {arXiv:2203.15668 [gr-qc]} \BibitemShut
  {NoStop}%
\bibitem [{\citenamefont {Yeh}\ \emph {et~al.}(2022)\citenamefont {Yeh},
  \citenamefont {Shelton}, \citenamefont {Olive},\ and\ \citenamefont
  {Fields}}]{Yeh:2022heq}%
  \BibitemOpen
  \bibfield  {author} {\bibinfo {author} {\bibfnamefont {T.-H.}\ \bibnamefont
  {Yeh}}, \bibinfo {author} {\bibfnamefont {J.}~\bibnamefont {Shelton}},
  \bibinfo {author} {\bibfnamefont {K.~A.}\ \bibnamefont {Olive}},\ and\
  \bibinfo {author} {\bibfnamefont {B.~D.}\ \bibnamefont {Fields}},\ }\href
  {https://doi.org/10.1088/1475-7516/2022/10/046} {\bibfield  {journal}
  {\bibinfo  {journal} {JCAP}\ }\textbf {\bibinfo {volume} {10}},\ \bibinfo
  {pages} {046}},\ \Eprint {https://arxiv.org/abs/2207.13133} {arXiv:2207.13133
  [astro-ph.CO]} \BibitemShut {NoStop}%
\bibitem [{\citenamefont {Armitage-Caplan}\ \emph {et~al.}(2011)\citenamefont
  {Armitage-Caplan}, \citenamefont {Avillez}, \citenamefont {Barbosa},
  \citenamefont {Banday}, \citenamefont {Bartolo}, \citenamefont {Battye},
  \citenamefont {Bernard}, \citenamefont {de~Bernardis}, \citenamefont {Basak},
  \citenamefont {Bersanelli} \emph {et~al.}}]{armitage2011core}%
  \BibitemOpen
  \bibfield  {author} {\bibinfo {author} {\bibfnamefont {C.}~\bibnamefont
  {Armitage-Caplan}}, \bibinfo {author} {\bibfnamefont {M.}~\bibnamefont
  {Avillez}}, \bibinfo {author} {\bibfnamefont {D.}~\bibnamefont {Barbosa}},
  \bibinfo {author} {\bibfnamefont {A.}~\bibnamefont {Banday}}, \bibinfo
  {author} {\bibfnamefont {N.}~\bibnamefont {Bartolo}}, \bibinfo {author}
  {\bibfnamefont {R.}~\bibnamefont {Battye}}, \bibinfo {author} {\bibfnamefont
  {J.}~\bibnamefont {Bernard}}, \bibinfo {author} {\bibfnamefont
  {P.}~\bibnamefont {de~Bernardis}}, \bibinfo {author} {\bibfnamefont
  {S.}~\bibnamefont {Basak}}, \bibinfo {author} {\bibfnamefont
  {M.}~\bibnamefont {Bersanelli}}, \emph {et~al.},\ }\href@noop {} {\bibfield
  {journal} {\bibinfo  {journal} {arXiv preprint arXiv:1102.2181}\ } (\bibinfo
  {year} {2011})}\BibitemShut {NoStop}%
\bibitem [{\citenamefont {Laureijs}\ \emph {et~al.}(2011)\citenamefont
  {Laureijs}, \citenamefont {Amiaux}, \citenamefont {Arduini}, \citenamefont
  {Augueres}, \citenamefont {Brinchmann}, \citenamefont {Cole}, \citenamefont
  {Cropper}, \citenamefont {Dabin}, \citenamefont {Duvet}, \citenamefont
  {Ealet} \emph {et~al.}}]{laureijs2011euclid}%
  \BibitemOpen
  \bibfield  {author} {\bibinfo {author} {\bibfnamefont {R.}~\bibnamefont
  {Laureijs}}, \bibinfo {author} {\bibfnamefont {J.}~\bibnamefont {Amiaux}},
  \bibinfo {author} {\bibfnamefont {S.}~\bibnamefont {Arduini}}, \bibinfo
  {author} {\bibfnamefont {J.-L.}\ \bibnamefont {Augueres}}, \bibinfo {author}
  {\bibfnamefont {J.}~\bibnamefont {Brinchmann}}, \bibinfo {author}
  {\bibfnamefont {R.}~\bibnamefont {Cole}}, \bibinfo {author} {\bibfnamefont
  {M.}~\bibnamefont {Cropper}}, \bibinfo {author} {\bibfnamefont
  {C.}~\bibnamefont {Dabin}}, \bibinfo {author} {\bibfnamefont
  {L.}~\bibnamefont {Duvet}}, \bibinfo {author} {\bibfnamefont
  {A.}~\bibnamefont {Ealet}}, \emph {et~al.},\ }\href@noop {} {\bibfield
  {journal} {\bibinfo  {journal} {arXiv preprint arXiv:1110.3193}\ } (\bibinfo
  {year} {2011})}\BibitemShut {NoStop}%
\bibitem [{\citenamefont {Ghiglieri}\ and\ \citenamefont
  {Laine}(2015)}]{Ghiglieri:2015nfa}%
  \BibitemOpen
  \bibfield  {author} {\bibinfo {author} {\bibfnamefont {J.}~\bibnamefont
  {Ghiglieri}}\ and\ \bibinfo {author} {\bibfnamefont {M.}~\bibnamefont
  {Laine}},\ }\href {https://doi.org/10.1088/1475-7516/2015/07/022} {\bibfield
  {journal} {\bibinfo  {journal} {JCAP}\ }\textbf {\bibinfo {volume} {07}},\
  \bibinfo {pages} {022}},\ \Eprint {https://arxiv.org/abs/1504.02569}
  {arXiv:1504.02569 [hep-ph]} \BibitemShut {NoStop}%
\bibitem [{\citenamefont {Ghiglieri}\ \emph {et~al.}(2020)\citenamefont
  {Ghiglieri}, \citenamefont {Jackson}, \citenamefont {Laine},\ and\
  \citenamefont {Zhu}}]{Ghiglieri:2020mhm}%
  \BibitemOpen
  \bibfield  {author} {\bibinfo {author} {\bibfnamefont {J.}~\bibnamefont
  {Ghiglieri}}, \bibinfo {author} {\bibfnamefont {G.}~\bibnamefont {Jackson}},
  \bibinfo {author} {\bibfnamefont {M.}~\bibnamefont {Laine}},\ and\ \bibinfo
  {author} {\bibfnamefont {Y.}~\bibnamefont {Zhu}},\ }\href
  {https://doi.org/10.1007/JHEP07(2020)092} {\bibfield  {journal} {\bibinfo
  {journal} {JHEP}\ }\textbf {\bibinfo {volume} {07}},\ \bibinfo {pages}
  {092}},\ \Eprint {https://arxiv.org/abs/2004.11392} {arXiv:2004.11392
  [hep-ph]} \BibitemShut {NoStop}%
\bibitem [{\citenamefont {Ringwald}\ \emph {et~al.}(2021)\citenamefont
  {Ringwald}, \citenamefont {Sch{\"u}tte-Engel},\ and\ \citenamefont
  {Tamarit}}]{Ringwald:2020ist}%
  \BibitemOpen
  \bibfield  {author} {\bibinfo {author} {\bibfnamefont {A.}~\bibnamefont
  {Ringwald}}, \bibinfo {author} {\bibfnamefont {J.}~\bibnamefont
  {Sch{\"u}tte-Engel}},\ and\ \bibinfo {author} {\bibfnamefont
  {C.}~\bibnamefont {Tamarit}},\ }\href
  {https://doi.org/10.1088/1475-7516/2021/03/054} {\bibfield  {journal}
  {\bibinfo  {journal} {JCAP}\ }\textbf {\bibinfo {volume} {03}},\ \bibinfo
  {pages} {054}},\ \Eprint {https://arxiv.org/abs/2011.04731} {arXiv:2011.04731
  [hep-ph]} \BibitemShut {NoStop}%
\bibitem [{\citenamefont {Drewes}\ \emph {et~al.}(2024)\citenamefont {Drewes},
  \citenamefont {Georis}, \citenamefont {Klaric},\ and\ \citenamefont
  {Klose}}]{Drewes:2023oxg}%
  \BibitemOpen
  \bibfield  {author} {\bibinfo {author} {\bibfnamefont {M.}~\bibnamefont
  {Drewes}}, \bibinfo {author} {\bibfnamefont {Y.}~\bibnamefont {Georis}},
  \bibinfo {author} {\bibfnamefont {J.}~\bibnamefont {Klaric}},\ and\ \bibinfo
  {author} {\bibfnamefont {P.}~\bibnamefont {Klose}},\ }\href
  {https://doi.org/10.1088/1475-7516/2024/06/073} {\bibfield  {journal}
  {\bibinfo  {journal} {JCAP}\ }\textbf {\bibinfo {volume} {06}},\ \bibinfo
  {pages} {073}},\ \Eprint {https://arxiv.org/abs/2312.13855} {arXiv:2312.13855
  [hep-ph]} \BibitemShut {NoStop}%
\bibitem [{\citenamefont {Konar}\ and\ \citenamefont
  {Show}(2025)}]{Konar:2025gvh}%
  \BibitemOpen
  \bibfield  {author} {\bibinfo {author} {\bibfnamefont {P.}~\bibnamefont
  {Konar}}\ and\ \bibinfo {author} {\bibfnamefont {S.}~\bibnamefont {Show}},\
  }\href@noop {} {\  (\bibinfo {year} {2025})},\ \Eprint
  {https://arxiv.org/abs/2512.13799} {arXiv:2512.13799 [hep-ph]} \BibitemShut
  {NoStop}%
\bibitem [{\citenamefont {Tito~D'Agnolo}\ and\ \citenamefont
  {Ellis}(2025)}]{TitoDAgnolo:2024res}%
  \BibitemOpen
  \bibfield  {author} {\bibinfo {author} {\bibfnamefont {R.}~\bibnamefont
  {Tito~D'Agnolo}}\ and\ \bibinfo {author} {\bibfnamefont {S.~A.~R.}\
  \bibnamefont {Ellis}},\ }\href {https://doi.org/10.1007/JHEP04(2025)164}
  {\bibfield  {journal} {\bibinfo  {journal} {JHEP}\ }\textbf {\bibinfo
  {volume} {04}},\ \bibinfo {pages} {164}},\ \Eprint
  {https://arxiv.org/abs/2412.17897} {arXiv:2412.17897 [gr-qc]} \BibitemShut
  {NoStop}%
\bibitem [{\citenamefont {He}\ \emph {et~al.}(2024)\citenamefont {He},
  \citenamefont {Giri}, \citenamefont {Sharma}, \citenamefont {Mtchedlidze},\
  and\ \citenamefont {Georgiev}}]{He:2023xoh}%
  \BibitemOpen
  \bibfield  {author} {\bibinfo {author} {\bibfnamefont {Y.}~\bibnamefont
  {He}}, \bibinfo {author} {\bibfnamefont {S.~K.}\ \bibnamefont {Giri}},
  \bibinfo {author} {\bibfnamefont {R.}~\bibnamefont {Sharma}}, \bibinfo
  {author} {\bibfnamefont {S.}~\bibnamefont {Mtchedlidze}},\ and\ \bibinfo
  {author} {\bibfnamefont {I.}~\bibnamefont {Georgiev}},\ }\href
  {https://doi.org/10.1088/1475-7516/2024/05/051} {\bibfield  {journal}
  {\bibinfo  {journal} {JCAP}\ }\textbf {\bibinfo {volume} {05}},\ \bibinfo
  {pages} {051}},\ \Eprint {https://arxiv.org/abs/2312.17636} {arXiv:2312.17636
  [astro-ph.CO]} \BibitemShut {NoStop}%
\bibitem [{\citenamefont {Domcke}\ \emph {et~al.}(2022)\citenamefont {Domcke},
  \citenamefont {Garcia-Cely},\ and\ \citenamefont {Rodd}}]{Domcke:2022rgu}%
  \BibitemOpen
  \bibfield  {author} {\bibinfo {author} {\bibfnamefont {V.}~\bibnamefont
  {Domcke}}, \bibinfo {author} {\bibfnamefont {C.}~\bibnamefont
  {Garcia-Cely}},\ and\ \bibinfo {author} {\bibfnamefont {N.~L.}\ \bibnamefont
  {Rodd}},\ }\href {https://doi.org/10.1103/PhysRevLett.129.041101} {\bibfield
  {journal} {\bibinfo  {journal} {Phys. Rev. Lett.}\ }\textbf {\bibinfo
  {volume} {129}},\ \bibinfo {pages} {041101} (\bibinfo {year} {2022})},\
  \Eprint {https://arxiv.org/abs/2202.00695} {arXiv:2202.00695 [hep-ph]}
  \BibitemShut {NoStop}%
\end{thebibliography}%

		
\

\clearpage
\newpage
\newpage
\maketitle
\onecolumngrid
\begin{center}
	\textbf{\large Unraveling Freeze-in Dark matter through the echoes of gravitational waves} \\ 
	\vspace{0.05in}
	{ \it \Large |Supplemental Material|}\\ 
	\vspace{0.05in}
	{Partha Konar$^1$ and Sudipta Show$^2$}\\
	\vspace{0.05in}
	$^1$ \sl{Theoretical Physics Division, Physical Research Laboratory, Ahmedabad, 380009, India}\\
	$^2$ \sl{Department of Physics, Indian Institute of Technology Kanpur, Kanpur 208016, India}
\end{center}
\onecolumngrid
\setcounter{table}{0}
\setcounter{section}{0}
\setcounter{page}{1}
\makeatletter

\section{Realization of different class of models}
\begin{table}[htb!] 
	\small
	\centering
	\begin{tabular}{|l|c|c|c|}
		\hline
		Model & $X$~(Mediator)	& $\xi$~(SM) & $\chi~$(DM)	\\    \hline  \hline
		I. & Vector-like & SM	& Scalar	\\    
		& quark & quark 	& \\    \hline
		II. & $\text{Vector-like}$ &  SM	& Scalar	\\    
		& lepton & lepton	& \\    \hline 
		
		III. & $SU(2)_L$ fermion & SM & Fermion  \\     
		& doublet &  Higgs &  \\     \hline
		
		IV. & Pair of $SU(2)_L$ Weyl & SM & Fermion  \\     
		& fermion doublets &  Higgs &  \\     \hline
	\end{tabular}
	\caption{Particles involved in the interaction. }
	\label{tab:Models}
\end{table}

The specific nature of DM defines the characteristics of the BSM and SM particles involved in its interactions. We summarize some of the attractive models based on the $y X\xi\chi$ interaction topology in TABLE~\ref{tab:Models}. Models I and II are commonly recognized as hadrophilic~\cite{Belanger:2018sti, Garny:2018icg, DEramo:2017ecx, Ghosh:2024nkj, Becker:2023tvd} and leptophilic DM models~\cite{Belanger:2018sti, Chakraborti:2019ohe, DEramo:2017ecx, Becker:2023tvd}, respectively. Conversely, Models III and IV illustrate singlet-doublet Dirac~\cite{No:2019gvl, Ghosh:2021wrk, Das:2023owa} and Majorana DM models~\cite{Calibbi:2015nha, Calibbi:2018fqf}.

\section{Evolution of DM and heavy particle number density}
\begin{figure}[htb!]
	\centering
	\includegraphics[width=8cm, height=6cm]{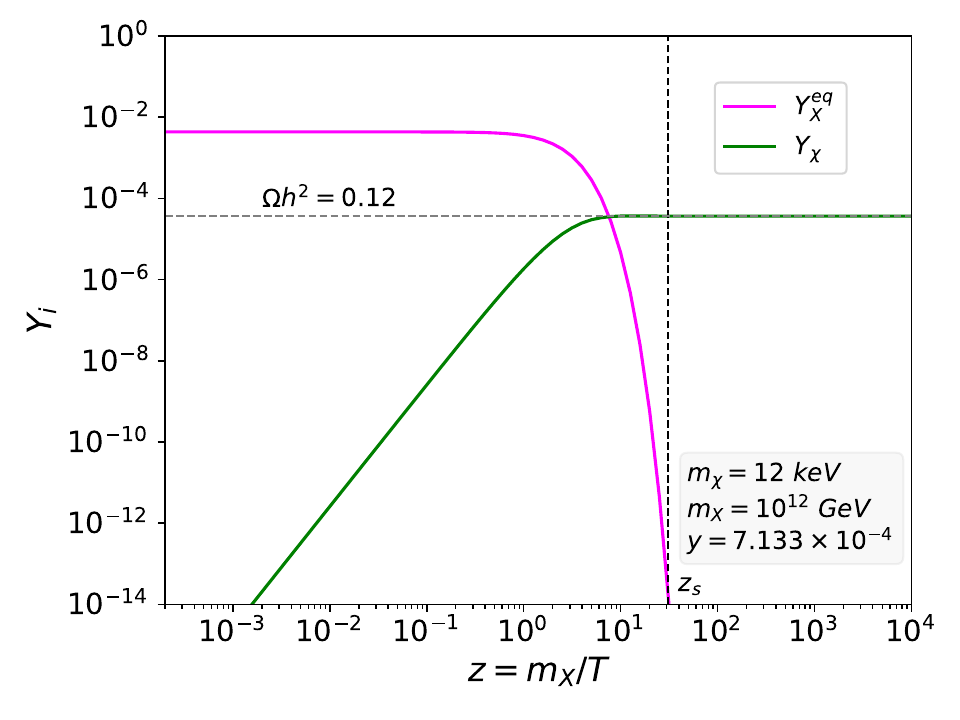}
	\caption{Evolution of $Y_X^{\text{eq}}$ and $Y_\chi$ as a function of the dimensionless variable $z=m_X/T$.}%
\label{abundance}
\end{figure}
For illustration purpose, in Figure~\ref{abundance}, we display both the evolution of the equilibrium abundance of $X$ ($Y_X^{\text{eq}}$) and abundance of DM ($Y_\chi$) as a function of $x$ for a given set of parameters $(m_\chi, m_X, y)$. Horizontal dashed line indicates the required abundance to satisfy the relic density constraint~\cite{Planck:2018vyg} for this benchmark.
Starting from zero abundance, DM generation continues until the number density of the decaying particles is significantly diluted due to Boltzmann suppression. Subsequently, the DM abundance saturates at a point denoted by $z_s$. 
When solving Eq.~(\ref{Boltz_abun}) for the superheavy mediator $(X)$, it's important to exercise caution regarding the initial value of \(z\), denoted as $z_{\text{in}}$, as it cannot be arbitrarily small. For instance, if we assume the reheat temperature\footnote{We focus on the maximum allowed reheating temperature for instantaneous reheating and plan to address the impact of reheating temperature on our findings in future research. However, reheating can occur over an extended period of time. In such a non-instantaneous reheating case, the temperature can drop to as low as 4 MeV. Interestingly, the production of dark matter in this era got modified.}, $T_{\text{rh}} (=5\times 10^{15}$ GeV), represents the maximum temperature of the universe, then the initial value $z_{\text{in}} (= m_X / T_{\text{rh}})$ should be $2\times10^{-4}$ (1) for $m_X = 10^{12}$ GeV ($5\times 10^{15}$~GeV).

\section{Non-thermalization of freeze-in dark matter}
\begin{figure}[htb!]
	\centering
	\includegraphics[width=8cm, height=6cm]{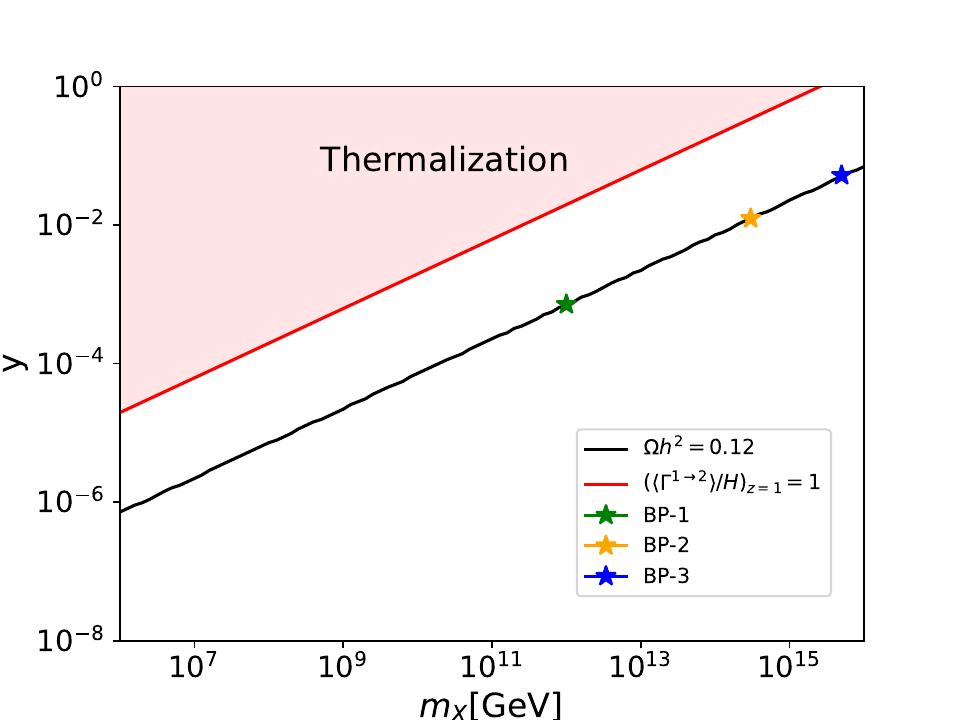}
	\caption{Plot shows the relic density satisfied contour and the parameter space disallowed by thermalization constraint.}
	\label{fig:ther}
\end{figure}
In a freeze-in scenario, it is essential to ensure that the dark matter candidate remains out of thermal equilibrium at all times. It is important to verify this, especially since the values of the Yukawa coupling in benchmarks are significantly higher than those typically seen in conventional freeze-in dark matter scenarios.
In Figure~\ref{fig:ther}, we illustrate the thermalization contour, represented by the red solid line, within the Yukawa coupling and mediator mass plane. This contour is determined by setting the ratio of the average decay rate, $\langle\Gamma\rangle$/H=1. To obtain a conservative bound, we evaluate this ratio at $z=1$. Therefore, the region above this contour is excluded due to thermalization. 
Additionally, we indicate the region that satisfies the required relic density in the same figure, marked by a black solid line. All the benchmarks used in our study are represented by stars, residing on this line. It is clear from the figure that dark matter candidates satisfy the relic density requirement across all benchmarks and also remain out of thermal equilibrium.

\section{Uniqueness of peak frequency of the GW spectrum}
It can be shown that the gravitational wave relic density is proportional to $\mathcal{F}(x)$; in particular,  $\Omega_{GW}h^2\propto x^2\mathcal{F}(x)$ where $x=E_{GW}/m_X$. It is important to note that the ultraviolet (UV) cut-off or the peak frequency of the gravitational wave spectrum can be obtained by maximizing the function  $x^2\mathcal{F}(x)$ which will provide the value of $x$($=x_{peak}$) that lies between 0 and 1/2 for the peak frequency. 
Furthermore, it is crucial to emphasize that the energy of the graviton ($E_{GW}$), which is used to define $x$, corresponds to the energy of the graviton at the time it was generated in the early universe. However, in Figure 2 of our manuscript, we present the gravitational wave spectrum as a function of present-day frequency. To calculate this, one must consider the redshifted energy.
So, the cut-off or peak frequency is given by
\begin{align}
	f^{peak}=\frac{E_{GW}^0	}{2\pi}=\frac{E_{GW}	}{2\pi}\frac{a_s}{a_0}=\frac{E_{GW}	}{2\pi}\frac{T_0}{T_s}
\end{align}
where the quantity with the superscript $0$ and without the superscript represent their values at today and when the GW production stopped in the early universe. We have neglected the $g_*$ dependence when going from the scale factor to the temperature variable. Here, $T_s$ represents the temperature when freeze-in production of dark matter saturates and the generation of gravitational waves stops. Now, to obtain the peak frequency, $E_{GW}$ can be replaced by $x_{peak}\times m_X$ and $T_s$ can be expressed in terms of $z_s$ and $m_X$. With this substitution, the expression of peak frequency becomes
\begin{align}
	f^{peak}=\frac{x_{peak} m_X}{2\pi}\bigg(\frac{T_0 z_s}{m_X}\bigg)=\frac{x_{peak} T_0 z_s}{2\pi}
\end{align}
It is clear from the above expression that the peak frequency is not only independent of $m_X$ but also constant, as all the parameters have fixed values. 

\section{Discussion on detection prospects of the GW spectrum}\label{Appen_3}
The cosmic gravitational wave background (CWWB) generated by the SM is consistently present, and it can mask any signal from new physics (NP), particularly in the high-frequency range where the proposed resonant cavity experiment is expected to operate. However, with a comprehensive understanding of the CGWB both theoretically and experimentally, it is still possible to explore such a signal. For instance, if we examine the first vertical band of cavity experiments at the frequency around $\sim 3\times10^7$ Hz, the GW amplitudes are expected at around $\sim2\times10^{-18}$ and $\sim2\times10^{-20}$ for CGMB and freeze-in associated GW, respectively. Therefore, it is feasible for such an experiment to indicate the excess contribution from the NP origin if it can measure the GW relic with (0.1 - 1) percent accuracy.

A clearer understanding of the GW signal generated by freeze-in can be conceived by employing a multi-frequency-band exploration across the spectrum, particularly focusing on the low-frequency tail of our GW spectra, where the CGWB contribution is minuscule. Current experiments in these bands, such as LIGO and the proposed uDECIGO, are unable to detect this low-frequency tail. 
With recent advances in gravitational wave detection techniques, significant enhancements in our capabilities and sensitivity are expected in this mid- to high-frequency range.
For instance, next-generation projects like the Einstein Telescope~\cite{Hild:2008ng} are developing cryogenic silicon mirrors that will significantly reduce thermal noise, allowing better sensitivity at the higher end of the kHz spectrum. In ref.~\cite{TitoDAgnolo:2024res}, the authors suggest that significant improvements in the sensitivity of laser interferometers can be achieved by leveraging advanced quantum sensing techniques in the future. A new frontier of Ultra-High-Frequency (UHF) detection is also emerging, targeting frequencies from kHz to GHz.
Such enhancements may enable detection of the mid-to-high-frequency distribution of the GW spectrum from heavy-particle bremsstrahlung.

\end{document}